\documentclass[10pt,journal,compsoc]{IEEEtran}
\pdfoutput=1
\ifCLASSOPTIONcompsoc
  \usepackage[nocompress]{cite}
\else
  \usepackage{cite}
\fi

\usepackage{graphicx}                
\DeclareGraphicsExtensions{.pdf,.png,.jpg,.jpeg,.eps} 
\graphicspath{{figures/}{pictures/}{images/}{./}} 

\usepackage{microtype}                 
\PassOptionsToPackage{warn}{textcomp}  
\usepackage{textcomp}                  
\usepackage{mathptmx}                  
\usepackage{times}                     
\usepackage{cite}                      
\usepackage{tabu}                      
\usepackage{booktabs}                  
\usepackage{multirow}
\usepackage{url}
\usepackage{amssymb}
\usepackage{amsmath}
\usepackage{float}
\usepackage{wrapfig}
\usepackage{subfigure}
\usepackage{color}
\usepackage[ruled, vlined]{algorithm2e}

\newcommand*{\code}[1]{\texttt{#1}}

\usepackage{calc}
\newlength\myheight
\newlength\mydepth
\settototalheight\myheight{Xygp}
\newcommand*\inlinegraphics[1]{%
  \settototalheight\myheight{Xygp}%
  \settodepth\mydepth{Xygp}%
  \raisebox{-.08\baselineskip}{\includegraphics[height=\myheight]{#1}}%
}
\usepackage{paralist}
\let\itemize\compactitem
\let\enditemize\endcompactitem
\let\enumerate\compactenum
\let\endenumerate\endcompactenum

\usepackage[table,xcdraw]{xcolor}
\usepackage{stfloats}

\begin{document}
%
\title{Towards Efficient Visual Simplification of Computational Graphs in Deep Neural Networks}
%
%
%
%


\author{Rusheng Pan, Zhiyong Wang, Yating Wei, Han Gao, Gongchang Ou, Caleb Chen Cao, Jingli Xu, Tong Xu and Wei Chen
\IEEEcompsocitemizethanks{
  \IEEEcompsocthanksitem Rusheng Pan, Zhiyong Wang, Yating Wei, Jingli Xu, Tong Xu, and Wei Chen are with the Stake key Lab of CAD\&CG, Zhejiang University, Hangzhou 310027, China. 
  \protect\\Email: panrusheng@zju.edu.cn, zerowangzy@outlook.com, weiyating@zju.\ edu.cn, xu1220341948@gmail.com, \{xutong8, chenvis\}@zju.edu.cn
  \IEEEcompsocthanksitem Han Gao, Gongchang Ou, Caleb Chen Cao are with the Distributed Data Lab, Huawei Technologies Co., Ltd., Shenzhen 518129, China. 
  \protect\\Email: \{gaohan19, ougongchang, caleb.cao\}@huawei.com.
  \IEEEcompsocthanksitem Wei Chen is the corresponding author.
\IEEEcompsocthanksitem Rusheng Pan and Zhiyong Wang contribute to this paper identically.}
}

%
%

\markboth{Journal of \LaTeX\ Class Files,~Vol.~14, No.~8, August~2015}%
{Shell \MakeLowercase{\textit{et al.}}: Bare Demo of IEEEtran.cls for Computer Society Journals}
%



\IEEEtitleabstractindextext{%
\begin{abstract}
A computational graph in a deep neural network (DNN) denotes a specific data flow diagram (DFD) composed of many tensors and operators. Existing toolkits for visualizing computational graphs are not applicable when the structure is highly complicated and large-scale (e.g., BERT~\cite{devlin2019bert}). To address this problem, we propose leveraging a suite of visual simplification techniques, including a cycle-removing method, a module-based edge-pruning algorithm, and an isomorphic subgraph stacking strategy. We design and implement an interactive visualization system that is suitable for computational graphs with up to 10 thousand elements. Experimental results and usage scenarios demonstrate that our tool reduces 60\% elements on average and hence enhances the performance for recognizing and diagnosing DNN models. Our contributions are integrated into an open-source DNN visualization toolkit, namely, MindInsight~\cite{MindInsight}.
\end{abstract}

\begin{IEEEkeywords}
  Deep neural networks, computational graphs, graph visualization, graph layout, visual simplifications
\end{IEEEkeywords}}

\maketitle

\IEEEdisplaynontitleabstractindextext

%
\IEEEpeerreviewmaketitle

\IEEEraisesectionheading{\section{Introduction}\label{sec:introduction}}

\begin{figure*}[h]
    \centering
    \includegraphics[width=1.0\linewidth]{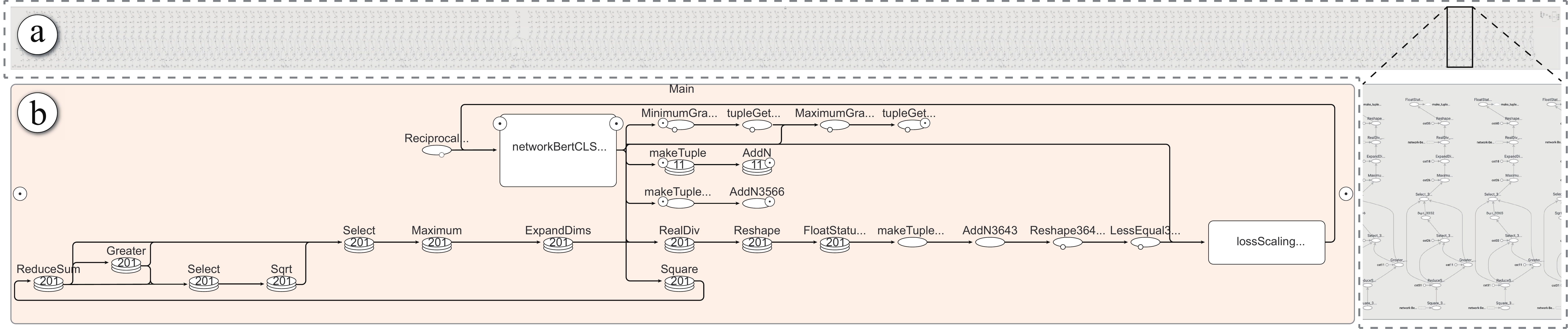}
    \caption{Visualizing the BERT model~\cite{devlin2019bert} by TensorBoard (a) and our approach (b). Its computational graph contains 5,698 nodes and 7,340 edges. Our result depicts sets of isomorphic subgraphs and meanwhile preserves the topological structures.}
    \label{fig:intro-bert}
    \vspace{-.2cm}
\end{figure*}

\IEEEPARstart{D}{eep} neural networks (DNNs) have been widely used in various fields, including computer vision, natural language processing, and autonomous driving~\cite{xu2018ecglens}. A DNN commonly uses a multilayer structure to learn and extract high-level features from the input dataset. Monitoring~\cite{ehrlinger2019daql}, understanding~\cite{du2018towards}, and modulating~\cite{kulesza2015principles} of models become challenging due to the complexity and abstractness of the associated DNN structures. To address these challenges, existing mainstream tools, such as TensorBoard~\cite{wongsuphasawat2017visualizing} and VisualDL~\cite{VisualDL}, use a visual interface and comprehensive widgets to support interactive training and applications of DNN models. A series of visual analysis tools for DNNs~\cite{DBLP:journals/tvcg/HohmanPRC20, DBLP:journals/tvcg/LiuSLLZL17, DBLP:conf/ieeevast/MingCZLCSQ17, DBLP:journals/tvcg/StrobeltGPR18, DBLP:journals/tvcg/WangGYS18} has been proposed by leveraging data flow diagram (DFD)~\cite{larman2012applying} to help developers understand and manipulate the data flow structure of DNNs. DNNs are represented as computational graphs for structural exploration and performance modulation. A computational graph is generally defined as a directed acyclic graph (DAG) and consists of tensors (e.g., activations and parameters) and operators (e.g., convolution and matrix multiplication). Tensors are represented as nodes or edges, and operators are normally represented as nodes. The layout, illustration, and exploration of computational graphs play a vital role in interpreting complex DNN models~\cite{samek2017explainable}.

Most efforts on visualizing DNN computational graphs~\cite{liu2017analyzing, kahng2017cti} have focused on optimizing the analysis process of model modulation. For instance, Net2Vis~\cite{bauerle2021net2vis} visualizes CNNs from Keras~\cite{chollet2015keras} code by automatically simplifying the data flow. Tools like draw\_convnet~\cite{draw_convnet} and convnet\_drawer~\cite{convnet_drawer} illustrate CNNs by automatically drawing diagrams of high-level structures of CNN layers. Their computational graph serves as an auxiliary for specific tasks, and lacks the scalability for general models. TensorBoard\cite{wongsuphasawat2017visualizing} and VisualDL~\cite{VisualDL} provide a general tool to visualize architectures of machine learning models.

Although these tools are usable in many situations, they are not efficient adequately to visualize large-scale computational graphs and the complex DNN structures of current deep learning (DL) models. On the basis of traditional DFD approaches, they suffer from visual clutter and performance bottlenecks. Without appropriate simplifications, visualizations have heavy visual clutter, significantly lowering the comprehensibility of DNNs. In addition, user interactions are difficult to perform. Classical focus+context techniques are no longer efficient and high-performance. For instance, BERT~\cite{devlin2019bert} contains a deep network hierarchy with 5368 tensors and operations. The user interface of TensorBoard becomes clunky when interacting with the computational graph of BERT because a number of visual elements need to be rendered (\figurename~\ref{fig:intro-bert}a). Consequently, it makes sense to achieve a fair balance between overview and details of exploring computational graphs of DNNs~\cite{devlin2019bert, krizhevsky2012alexnet, lecun1998lenet, sandler2018mobilenetv2, he2016resnet}, which contain plenty of elements and complex structures. 

This work focuses on hierarchical clustered computational graphs~\cite{jia2019optimizing}, whose visualization for general exploration scenes remains challenging: 

\textbf{C1.} Nodes in a computational graph of a DNN have deep hierarchies. An overview of a large-scale computational graph is too complex to explore. Abstracting the structures from different hierarchies remains challenging.

\textbf{C2.} Computational graphs often contain many tensors and operators whose visualizations exhibit heavy visual clutter, such as edge crossings and high node density. Conventional representation approaches can hardly handle their complex structures and a wide variety of DNNs. 

\textbf{C3.} Computational graphs have many similar or isomorphic structures. Manipulating these clustered structures is visually unpleasing (e.g., \figurename~\ref{fig:intro-bert}a). Achieving a fair balance between overview (e.g., highlighting essential graph structures) and details (e.g., maintaining the topological relationships) brings challenges. 

To address these challenges, we propose a visual simplification approach with a suite of intuitive visual analysis techniques for computational graphs. 
Our contributions include a cycle-removing method that effectively detects and removes directed cycles generated by the hierarchical structures, a module-based edge-pruning algorithm that can retain the salient information of computational graphs, and an isomorphic subgraph stacking technique that can automatically merge similar or identical graph substructures. We propose an integrated workflow and implement a prototype system based on the open-source toolkit MindSpore~\cite{MindSpore}.

We examine the efficiency of our solution with representative models like: BERT~\cite{devlin2019bert} and ResNet-50~\cite{he2016resnet}. Experimental results indicate that our implementation can effectively handle computational graphs with up to 10 thousand elements. Result comparisons verify that the proposed approach outperforms TensorBoard in terms of performance, visual {complexity}, and user interactions. \figurename~\ref{fig:intro-bert}b shows our result for BERT, which exhibits a more compact and straightforward structure than that of TensorBoard (\figurename~\ref{fig:intro-bert}a). Moreover, we conduct a user study and interview experts to evaluate the utility and effectiveness of our system. 

The rest of this paper is organized as follows. Section~\ref{sec:related-work} reviews related works, followed by the background description in Section~\ref{sec:background}. The requirements are analyzed in Section~\ref{sec:requirements}, and our approach is elaborated in Section~\ref{sec:algorithms}. Evaluation results and discussions are presented in Sections~\ref{sec:evaluation} and ~\ref{sec:discussion}. Section~\ref{sec:conclusion} concludes this work.

\section{Related Work}\label{sec:related-work}
This section presents the relevant studies on the visual explanation of DNNs, the computational graph drawing, and visual exploration of hierarchical graphs.
\subsection{Visual Explanation of DNNs}
Explainable artificial intelligence (XAI) aims to solve the problem of the ``black box" model in artificial intelligence~\cite{gunning2019darpa, LIU201748}. Research on reliable visualization tools to understand machine learning models has become prominent in the visualization community over the past decades~\cite{chatzimparmpas2020state, Du2019Tech}. As surveyed by Yuan et al. ~\cite{DBLP:journals/cvm/YuanCYLXL21}, existing efforts on visual analytics techniques for DNNs can be divided into three categories: techniques before~\cite{DBLP:journals/tvcg/KrausePB14, DBLP:journals/tvcg/BernardHZFS18}, during~\cite{DBLP:journals/tvcg/HohmanPRC20, DBLP:journals/tvcg/AhnL20, DBLP:journals/tvcg/DingenVHMKBW19}, and after~\cite{DBLP:journals/tvcg/BergerMS17, DBLP:journals/tvcg/AndrienkoAABBFH21} model building. Our work is designed to enhance the exploration and diagnosis of computational graphs, which belongs to the techniques after model building. 

XAI methods can be divided into two categories, namely, internal and external model interpretation~\cite{1532820}. The internal ones directly visualize the training process and structures to disclose the ``black box". Alternatively, the external ones use the input and output of the underlying model to train a new model to assist in understanding the model by simulating the ``black box". From this perspective, our work follows the category of the internal model interpretation by illustrating and analyzing the computational graph to uncover the data flow process and the model architecture.

TensorBoard~\cite{wongsuphasawat2017visualizing} graph visualizer excels at visualizing the computational graph of machine learning models and tracking various histograms and metrics of models and datasets. Although effective in many cases, TensorBoard can hardly handle complicated computational graphs. Our work tackles the key problem of simplifying computational graphs to improve the interpretability and comprehensibility of computational graphs.

\subsection{Computational Graph Drawing}\label{subsec:DFD}

To help uncover the underlying dependencies between data flow of the increasingly complicated neural networks. Tzeng et al.~\cite{1532820} propose to represent each neuron in the neural network as a node. Nowadays, computational graph drawing is a popular tool to depict the DFD of DNNs~\cite{DBLP:journals/corr/Owhadi2022}. 

Different approaches are adopted to represent operators and tensors of the computational graph. Activis~\cite{kahng2017cti} encodes operators and tensors as nodes with different shapes and uses edges to represent connections between pairs of a tensor and an operator. The generated bipartite graph provides an easy-tracking topological structure but contains masses of nodes. TensorBoard represents operators as nodes and tensors as edges, which provides a relatively succinct graph. However, it is challenging to understand, especially for DNNs with complicated forward propagation patterns or quantities of isomorphic subgraphs. We propose to stack graph elements to reduce the number of nodes while maintaining the comprehensibility of the computational graph.

Various layout styles are used to draw computational graphs. Most works and tools including TensorBoard, VisualDL~\cite{VisualDL} (whose computational graph module is powered by NETRON~\cite{NETRON}), and HiddenLayer~\cite{HiddenLayer} use a layered graph drawing algorithm, called Sugiyama Algorithm~\cite{sugiyama1981methods}. It first calculates the node layers from the downward directed edges and positions the nodes in orderly rows. Dummy nodes are introduced to represent the bends of edges that cross the layers during the entire layout process but are hidden in the final result. In each layer, nodes, including the dummy nodes, are sorted by using a heuristic algorithm to minimize edge crossings. During the ordering process, edges with bends at dummy nodes are determined. To fully use the display space, Activis~\cite{kahng2017cti}, DGMTracker~\cite{liu2017analyzing} and Net2Vis~\cite{bauerle2021net2vis} place the flow direction from left to right. VisualDL and HiddenLayer support both flow directions.  

Apart from the layer-based edge routing, the orthogonal layout \cite{6312901, Eschbach06orthogonalhypergraph} can be used to depict DFD. Specifically, Eades et al.~\cite{eades1996orthogonal} propose an orthogonal grid drawing of the clustered graphs, reducing crossing edges and improving the screen space utilization. The orthogonal hierarchical layout of DFD~\cite{DBLP:conf/gd/EadesF96, 221135, 10.1007/3-540-58950-3_371} can be enhanced by means of the port constraint~\cite{10.1007/978-3-642-11805-0_14}, where a port refers to the attachment points of the edges. Our work adopts the layered layout algorithm to generate an orthogonal style. 

\subsection{Visual exploration of hierarchical graphs}

The visual exploration of graphs plays a vital role in the field of visual analytics. For large and complex graphs, people usually view part or all of the graph at different levels of detail because the graphs can hardly be displayed at one time on a single screen~\cite{DBLP:journals/csur/CockburnKB08}. Thanks to the hierarchical clustering technique, where nodes are grouped into clusters by superimposing hierarchies on them, people can navigate the clusters until an appropriate level of the hierarchy is reached~\cite{DBLP:journals/tochi/SchafferZGBDDR96}. 

Generally, effective visual analysis of graphs requires appropriate visual presentations in combination with separate user interaction facilities and algorithmic graph analysis methods~\cite{von2011visual}. To view hierarchically clustered networks both with local detail and global context, many dedicated techniques are employed for visually simplified representation of clusters and edges. In addition to the static representation, many efforts focus on transition techniques for the continuously adaptive and adjustable views of the clustered graphs~\cite{DBLP:conf/apvis/BalzerD07}. 
Huang et al.~\cite{DBLP:journals/vlc/HuangL06} cluster graphs via node similarity and hierarchically draw the graphs in different abstract level views. Our computational graph data is generated with namespaces (see \figurename~\ref{fig:pipeline}a), which can be used to yield the graph hierarchy. However, the primitive hierarchies would make the semantic relationships of graphs misinterpreted. Appropriate adjustments to the namespaces are necessary.
FADE~\cite{DBLP:conf/gd/QuigleyE00} is a fast algorithm for multilevel viewing of large undirected graphs, which includes edges and multilevel visual abstraction. Our hierarchical directed graph is drawn by using ELK\footnote{https://www.eclipse.org/elk/} (Eclipse Layout Kernel, an open-source library of layered layout algorithms) recursively in each hierarchy (details are introduced in Section ~\ref{subsection:implementation}).

In our work, we propose a series of visual simplification techniques for nodes within a cluster and edges across different clusters, with the goal of relieving visual clutter of static layout and improving the fluency of transitions.

\section{Background}\label{sec:background}
This section briefly introduces related concepts of computational graphs of DNNs and MindSpore.

\begin{figure}[H]
    \vspace{-.2cm}
    \centering
    \includegraphics[width=.99\linewidth]{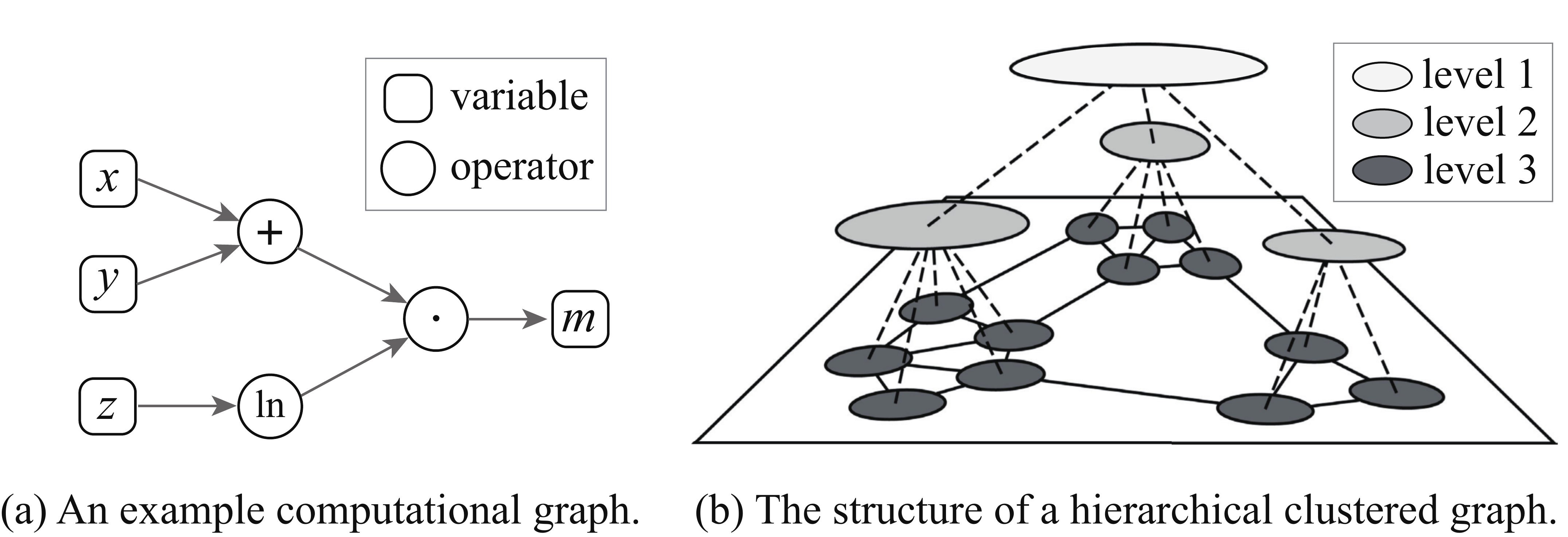}
    \vspace{-.2cm}
    \caption{Illustration of a computational graph and a hierarchically clustered graph.}
    \label{fig:computational-graph}
\end{figure}

\textbf{Computational Graph and DNN.}
A computational graph commonly represents a function based on the graph theory. For instance, \figurename~\ref{fig:computational-graph}a illustrates a function $m = f(x,y,z) = (x+y) \cdot \ln z$. 
Here, $f$ is a typical composite function: $f = g \cdot h$, where $g = x+y, h = \ln z$. A DNN is composed of numerous functions and trained by iteratively adjusting parameters of these functions~\cite{larochelle2009exploring}. An optimal combination of these parameters is computed to achieve the highest performance of the underlying DNN.

A computational graph is a DAG whose nodes represent tensors and operators, and the edges denote the data flow between them. It is visually overwhelming to display a holistic layout of the large-scale computational graph. Organizing computational graphs hierarchically makes sense because it provides a valuable level of abstraction for users~\cite{DBLP:journals/tvcg/ArchambaultMA08}. Namespaces are used by TensorFlow~\cite{abadi2016tensorflow} and MindSpore~\cite{MindSpore} to yield a hierarchically structured graph representation. The namespaces are generated automatically or in a user-defined manner.

The hierarchy of a graph is defined as a recursive grouping placed on the nodes of the initial graph~\cite{DBLP:journals/tvcg/ArchambaultMA08}. The nodes of a hierarchical graph contain metanodes and leaves. Metanodes denote subgraphs that contain a subset of nodes and a subset of the edges between these nodes (e.g., nodes of level 1 and 2 in \figurename~\ref{fig:computational-graph}b); leaves are the nodes of the input graph (e.g., nodes of level 3 in \figurename~\ref{fig:computational-graph}b). We represent a hierarchical computational graph as $H_0 = (H_1, E_1)$, where $ H_i= (H_{i+1}, E_{i+1}), E_{i} \subset H_i^2 (i=1,2,...,n)$. $H_i$ represents a set of nodes at the $i$-th level of the hierarchy; $E_{i}$ denotes a set of edges between $H_{i}$.

\textbf{Generating computational graphs} relies on the construction of the DNN models. Conventional approaches for model construction can be classified into two categories. The first category, e.g., TensorFlow~\cite{abadi2016tensorflow}, constructs a static graph before execution. It defines all operations and network structures. A static compilation technology optimizes the network performance, but it isn't very easy for model development and modulation. The second category, e.g., PyTorch~\cite{paszke2019pytorch}, performs dynamic graph calculation on the fly without defining the entire graph. It is more flexible and comfortable to modulate at the expense of performance. This process does not build any graph because the data flow graph is generated dynamically. The computational graph cannot easily represent the model structures.

\textbf{MindSpore} is an open-source framework~\cite{MindSpore} that allows creating new DNNs or leveraging built-in DNNs for fields, such as computer vision, natural language processing, and graph processing. Third-party framework models including TensorFlow and PyTorch can be converted into MindSpore. MindSpore provides an encoding mode to unify dynamic and static graphs, which optimize static compilation while efficiently providing flexible interfaces to build models. Similar to TensorBoard for TensorFlow, MindInsight~\cite{MindInsight} provides a suite of visualization widgets for MindSpore. As an open-source toolkit, MindInsight has already been used by actual users. Our work tackles the kernel of MindInsight that visualizes computational graphs of DNNs and contributes a general solution for visualizing the static computational graphs generated by MindSpore. The source code of our work is available in its latest repository\footnote{https://github.com/ZeroWangZY/DL-VIS}.

\section{Requirement Analysis}\label{sec:requirements}

To characterize the domain challenges and identify the analytical requirements, we conduct remote sessions weekly with experts of MindSpore (co-authors of this paper), gather feedback from MindInsight developers from the open source community, and summarize the experience during the implementation. We identify the following requirements.
\begin{itemize}

\item [\textbf{R1:}] \ \ \ \textbf{Clarify connective relationships with details on demand.} 
\item \textbf{R1.1:} \textbf{Clear and compact graph structures are necessary} from a global point of view. Large quantities of edges tend to intertwine and cross one another, confusing the connective relationships between nodes. Simplifying less critical elements relieves visual cluttering and ambiguity.
\item \textbf{R1.2:} \textbf{Users are allowed to focus on connection details of interest} in exploring the simplified graph. Interactions, such as disclosing primitive connections, will help users acquire accurate local details to understand the model architecture.

\item [\textbf{R2:}] \ \ \ \textbf{Simplify complex structures among relevant context.}
\item \textbf{R2.1:} \textbf{The rendering performance and readability} may decline when displaying a large number of isomorphic structures. It is highly required to represent the isomorphic structures with scalable glyphs.

\item \textbf{R2.2:} \textbf{Depicting relevant context surrounding simplified structures} is meaningful, and thereby lost details should be reminded to some degree. 

\item [\textbf{R3:}] \ \ \ \textbf{Maintain users' mental map during explorations.}
\item \textbf{R3.1:} \textbf{Maintaining a consistent transition of graph layout} is essential. For instance, when a metanode is expanded, its connected edges are divided into several line segments by the enlarged node boundary. Our approach should consider maintaining visual consistency by designing position constraints where metanodes and edges are connected.
\item \textbf{R3.2:} \textbf{The graph layout should conform to users' domain knowledge.}
The inherent forward propagation graph structure shows the neural network in a forward direction from the input layer to the output layer, and plays a core role in helping to understand the principles and parameters of the model. In principle, a computational graph is a DAG and does not contain cycles. Otherwise, the data flow would fall into an infinite loop and never end. However, hierarchical structures generated by grouping by the namespaces of MindSpore may produce directed cycles between metanodes (\figurename~\ref{fig:cycle} a and b). Consequently, an appropriate preprocessing should be performed to remove directed cycles before displaying the forward propagation graph structure.

\end{itemize}

\section{Our Approach}\label{sec:algorithms}
We propose a series of techniques for the visual simplification of DNN computational graphs. First, we define and represent the graph elements of the computational graph. Then, we elaborate the three approaches for computational graph layout. Finally, the user exploration and implementation of our computational graph visualization tool are introduced.

\subsection{Definitions and Representations}

\begin{figure}[ht]
    \centering
    \includegraphics[width=.9\linewidth]{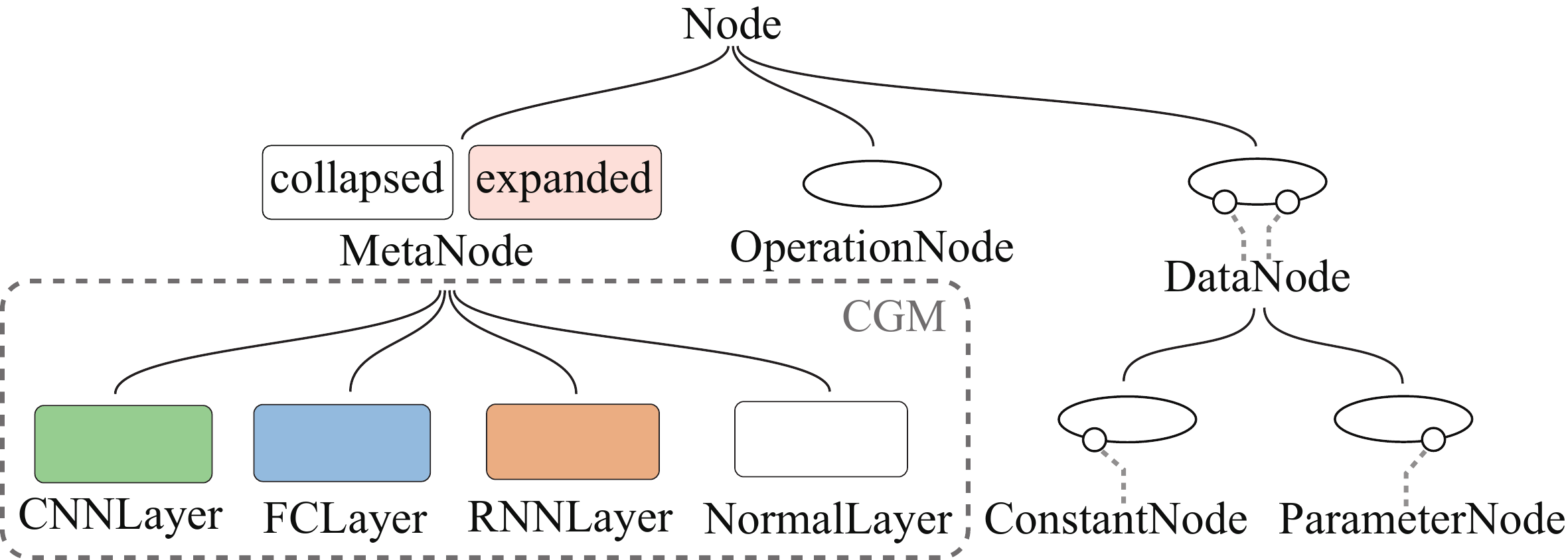}
    \vspace{-.1cm}
    \caption{The hierarchy and visual encoding of various types of nodes. }
    \label{fig:node-definition}
\end{figure}

We separate nodes into \code{OperationNode}, \code{DataNode}, and \code{MetaNode}. \code{OperationNode} refers to nodes of operators, and \code{DataNode} refers to nodes of tensors. \code{DataNode} can be classified as \code{ConstantNode} or \code{ParameterNode}. A \code{MetaNode} refers a subgraph. A \code{MetaNode} that contains many descendants (a user-defined threshold) can be further defined as a \code{ModuleNode}. Otherwise, it is called \code{NonModuleNode}. The detailed definition of ``module" is introduced in Section~\ref{subsubsec:module-based-edge-pruning}. To highlight the forward propagation graph structure, we provide a concept graph mode (CGM) (shown in \figurename~\ref{fig:system-interface}b). The CGM employs the cycle-removing algorithm (see Section~\ref{subsubsec:cycle-removing}) and abstracts \code{MetaNodes} into various DNN layers. In the CGM, \code{MetaNodes} are divided into \code{CNNLayer} (Convolutional Neural Network layer), \code{RNNLayer} (Recurrent Neural Network layer), \code{FCLayer} (Fully Connected layer), and \code{NormalLayer} (other layers). The hierarchy and representations of the nodes are illustrated in \figurename~\ref{fig:node-definition}.

\begin{figure}[H]
    \centering
    \includegraphics[width=.8\linewidth]{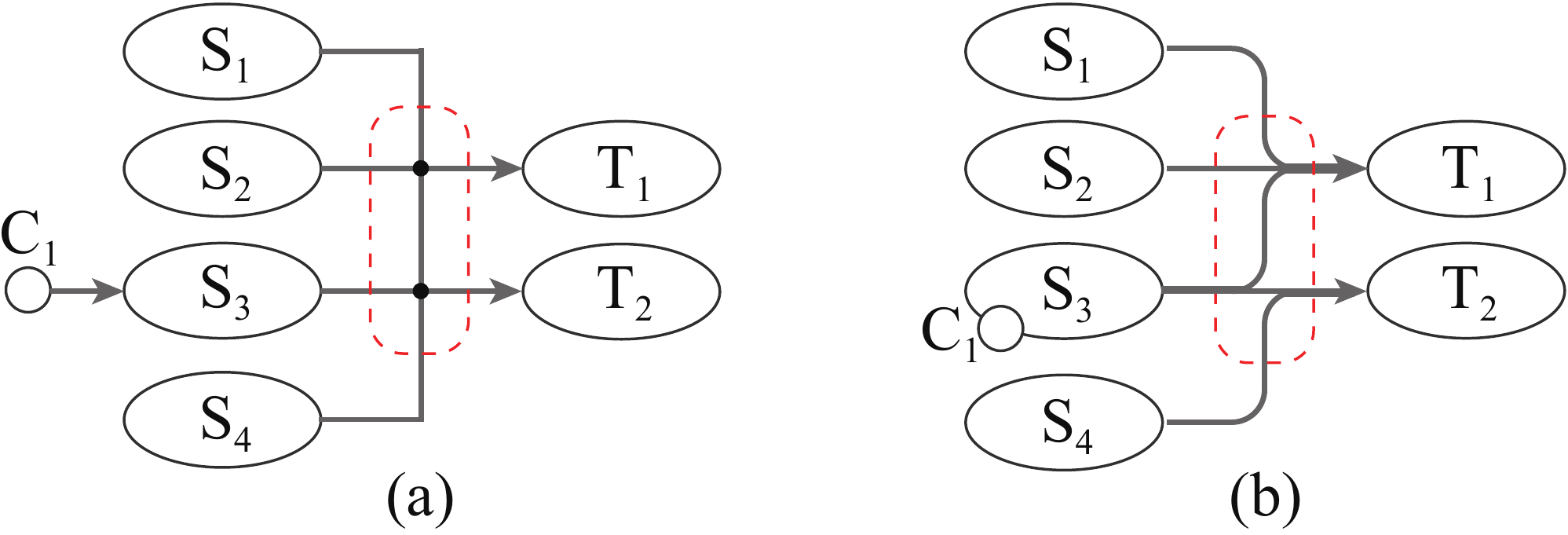}
    \vspace{-.2cm}
    \caption{The illustration of the orthogonal edge routing. (a) The right-angle transition at bends causes profound ambiguity. (b) After representing bends with a circular arc transition, edges are represented correctly.}
    \label{fig:edge-design}
\end{figure}

All edges of a computational graph are directed edges. The edges can be categorized as: \code{ModuleEdge}, \code{HiddenEdge} and \code{NormalEdge}, whose definitions and corresponding visual encodings are given in Section~\ref{subsubsec:module-based-edge-pruning}.
The edges are routed orthogonally to avoid edge crossing. The right-angle transition at the edge bends generates junction points (the red dashed box of \figurename~\ref{fig:edge-design}a). It causes ambiguity of the source and target of an edge. Therefore, we use a circular arc transition, as shown in \figurename~\ref{fig:edge-design}b. It clearly shows that nodes $S_1$, $S_2$, and $S_3$ link to node $T_1$, and $S_3$ and $S_4$ link to $T_2$. 
In DNN models, a \code{DataNode} always targets to an \code{OperationNode}. Consequently, edges between two \code{DataNodes} or between a \code{DataNode} and a \code{MetaNode} do not exist. For the general visual design of the layout, we attach a \code{DataNode} to an \code{OperationNode} to represent their connections because edges between a \code{DataNode} and an \code{OperationNode} are unidirectional. To be specific, we place a \code{ConstantNode} on the bottom left of an \code{OperationNode}, and a \code{ParameterNode} on the bottom right. For instance, the edge from \code{ConstantNode} $C_1$ to \code{OperationNode} $S_3$ (\figurename~\ref{fig:edge-design}a) is represented by attaching $C_1$ to the bottom left of $S_3$ (\figurename~\ref{fig:edge-design}b).

\begin{figure*}[h]
    \centering
    \includegraphics[width=1.0\linewidth]{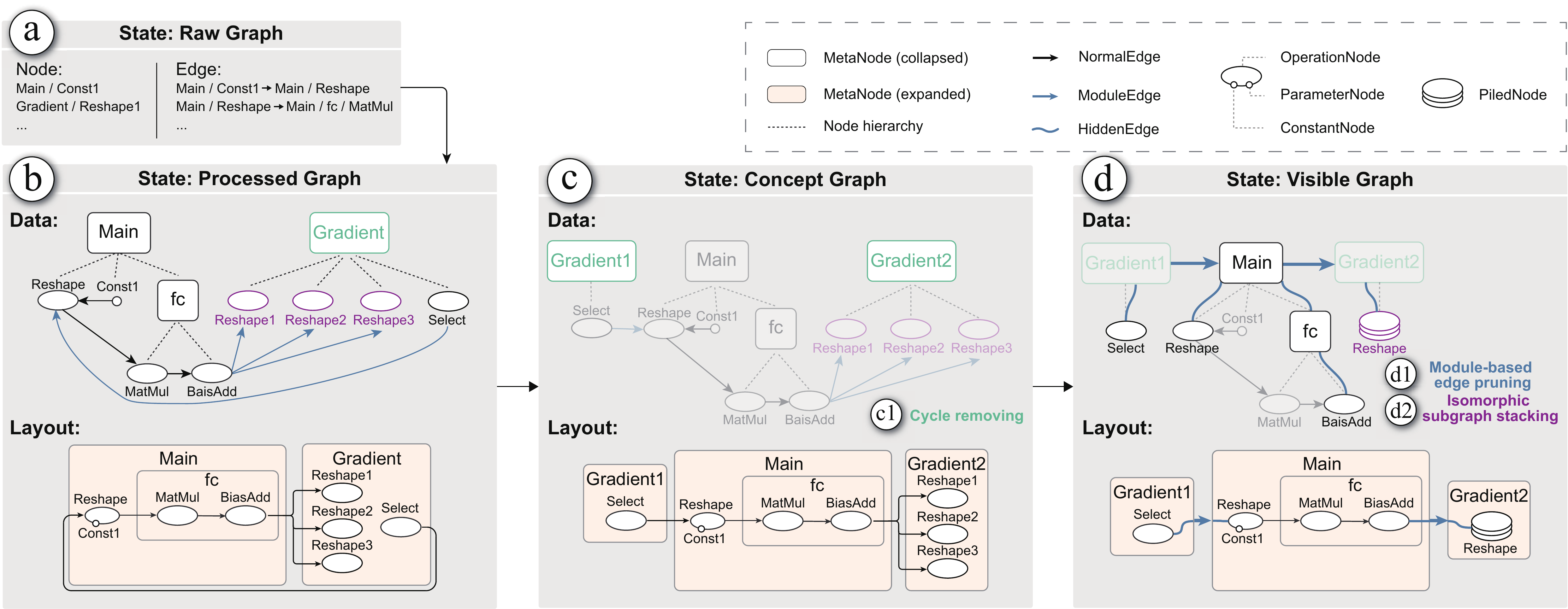}
    \caption{The overall workflow of our approach: different graph states and the corresponding layout. The components that are unchanged between two graph states are faded. (a) A raw graph is initialized as a computational graph by parsing the summary file. (b) A processed graph denotes the optimized graph. If the concept graph (c) mode is set, the cycle-removing scheme (c1) will be applied to the processed graph. (d) A visible graph is derived from the processed graph by the module-based edge pruning algorithm (d1) and isomorphic subgraph stacking scheme (d2). }
    \label{fig:pipeline}
    \vspace{-.3cm}
\end{figure*}

\subsection{Computational Graph Layout}
The layout workflow contains four graph states, their data structures and the corresponding graph layout are illustrated in \figurename~\ref{fig:pipeline}. 
\begin{itemize}
\item[1)] \textbf{Raw Graph}
denotes the computational graph structure data generated from the input DNN model by parsing the summary file. The raw graph contains nodes of different hierarchies and edges between nodes.
 
\item[2)] \textbf{Processed Graph}
is created from the raw graph data.
We organize the hierarchical structure information of nodes in a tree and identify the leaf nodes' edges.

\item[3)] \textbf{Concept Graph} 
applies a cycle-removal scheme (see Section~\ref{subsubsec:cycle-removing}) to the processed graph, which optimizes the hierarchical structures (R3.2). 
 
\item[4)] \textbf{Visible Graph}.
Only expanded nodes of the processed graph are visible in the expanded structures. A visible graph is generated by means of the module-based edge-pruning algorithm (R1, R3.1), which is introduced in Section~\ref{subsubsec:module-based-edge-pruning}. 
After that, an isomorphic subgraph stacking operation (R2) is performed to simplify the visible graph (Section~\ref{subsubsec:isomorphic-subgraph-stacking}).

\end{itemize}

We elaborate on three algorithms in the following subsections. 

\subsubsection{Cycle Removing}\label{subsubsec:cycle-removing}
\begin{figure}[H]
  \centering
  \includegraphics[width=.7\linewidth]{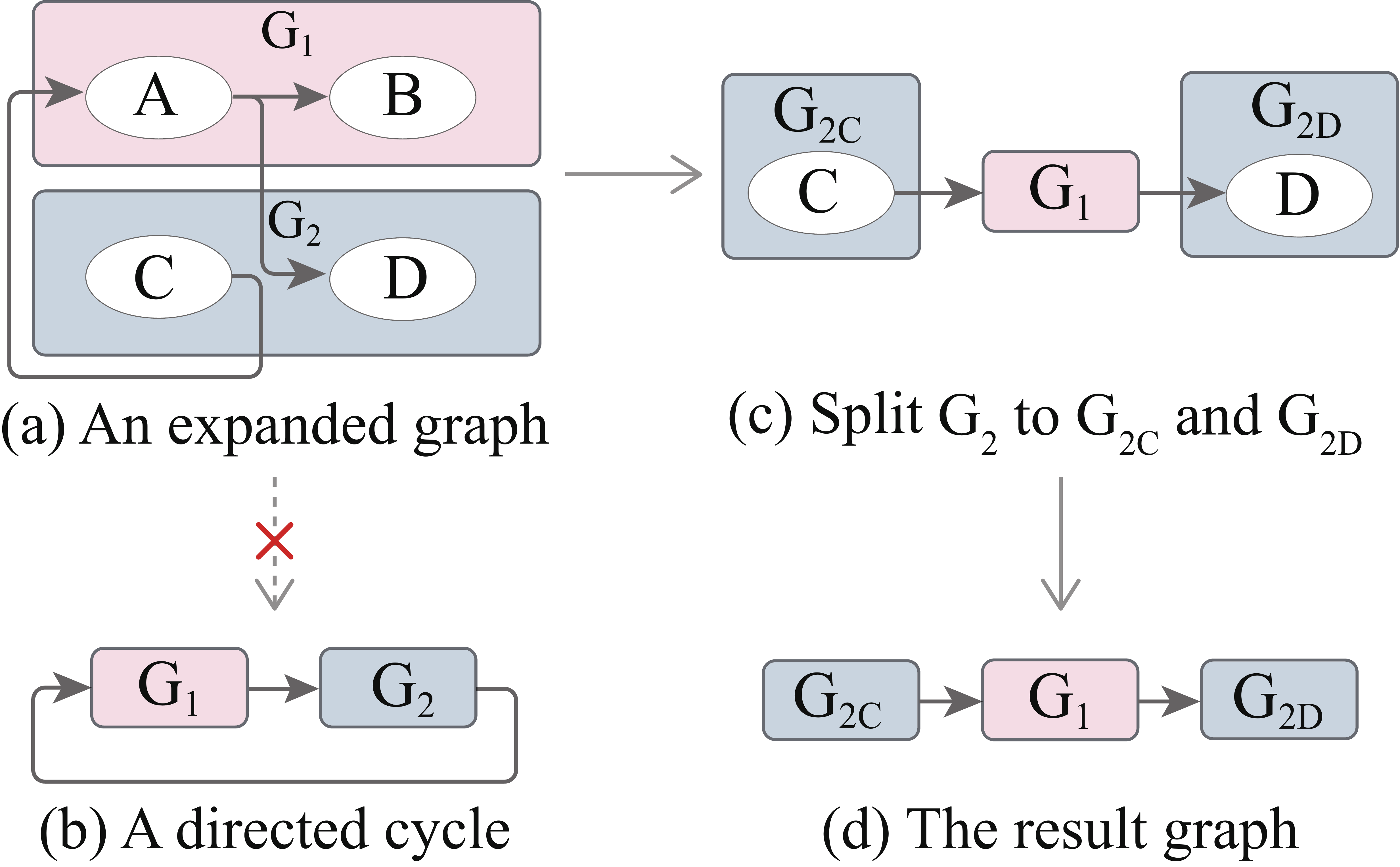}
  \vspace{-.2cm}
  \caption{(a) shows a computational graph after expanding $G_1$ and $G_2$. (b) The collapsed \code{Metanodes} $G_1$ and $G_2$ form a simple directed cycle. (c) Our algorithm splits $G_2$ into $G_{2C}$ (composed of $C$) and $G_{2D}$ (composed of $D$). (d) The result graph contains no cycle.}
  \label{fig:cycle}
  \vspace{-.1cm}
\end{figure}

A computational graph should be simplified to highlight the forward propagation graph structure (R3.2), whose tensors and operators do not contain directed cycles (``cycle" in this paper refers to ``directed cycle"). However, the processed graph may contain many ``false" cycles, misleading the data flow. These cycles could be generated when grouping the operators, which are not connected, into a metanode. The cycles could be composed of these metanodes. Consequently, the cycles make it difficult to understand the forward propagation graph. \figurename~\ref{fig:cycle}a and b illustrate the generation of the cycles. The graph (\figurename~\ref{fig:cycle}a) is composed of $A$, $B$, $C$, and $D$ does not have a cycle. After the grouping process of the DL framework, $A$ and $B$ are grouped into $G_1$; $C$ and $D$ are grouped into $G_2$ (\figurename~\ref{fig:cycle}a). This yields a cycle composed of $G_1$ and $G_2$.

In the CGM, we try to restore the forward propagation graph structure by removing the cycles. Simply expanding all related \code{MetaNodes} works but will dramatically increase the number of elements in the graph. Accordingly, we propose to remove them by ungrouping one of the \code{MetaNodes} in the cycle and rebuild two new \code{MetaNodes}. 

To detect all cycles, we start from a \code{DataNode} which has no input. We traverse all \code{OperationNodes} whose input is a \code{DataNode} because a \code{DataNode} attaches to an \code{OperationNode}. In traversing all nodes, we record the parent node of each visited node. If a node $n$ has no inclusion relationship with the last visited one, and $n$'s parent node has been recorded, then its parent node must be in a cycle. We remove the cycle by splitting its parent node, which is a \code{MetaNode}.

\vspace{-.1cm}

\figurename~\ref{fig:cycle} illustrates this process: (a) displays an expanded computational graph composed of $A$, $B$, $C$, and $D$, where no cycles exit. As shown in (b), collapsing $G_1$ and $G_2$ would generate a cycle of $G_1$ and $G_2$. To remove the cycle, we choose first to ungroup $G_2$, and group $C$ into $G_{2C}$ and group $D$ into $G_{2D}$ with the cycle-removing strategy. As a result, the generated graph in (c) contains no cycles. The pseudocode of our algorithm is shown in Appendix A. The time complexity of our cycle-removing algorithm is $O(|E|)$. This algorithm can remove almost all cycles in $H_0$ ($H_0$ represents all the nodes at the top level of the hierarchy). The detailed information is discussed in Section~\ref{sec:discussion}.



\subsubsection{Module-based Edge Pruning}\label{subsubsec:module-based-edge-pruning}
Classical edge bundling~\cite{holten2006hierarchical} algorithms can alleviate visual clutter to some extent and improve the entire topological structure conciseness~\cite{7192715}. However, bundling edges may cause the ambiguity of the nodes' connective relationships in the hierarchically clustered computational graph and fail to keep primitive details inside the \code{MetaNodes}, such as continuity of the primitive edges. 

We propose a module-based edge pruning algorithm that is specifically designed for large-scale computational graphs. It consists of two stages, namely, module recognition and edge update.

In the first stage, all the \code{MetaNodes} are traversed to yield the module information. A \code{MetaNode} would be identified as a \code{ModuleNode} or ``module" (R1.1), if its number of descendants is greater than the user-defined threshold. The module level of a \code{ModuleNode} is equivalent to its level of the hierarchy. In the next stage, we update all edges based on the identified modules. Edges across modules of different levels that share common line segments are considered for the update. Specifically, an edge from \code{OperationNode} $s$ to \code{OperationNode} $t$ can be sliced into at most three line segments, namely, the line segment from $s$ to $s$'s parent, the line segment from $s$'s parent to $t$'s parent, and the line segment from $t$'s parent to $t$. The detailed strategies are elaborated as the pseudocode in Appendix B. The time complexity is $O(|N|+|E|)$.  

\begin{figure}[h]
  \centering
  \includegraphics[width=1.0\linewidth]{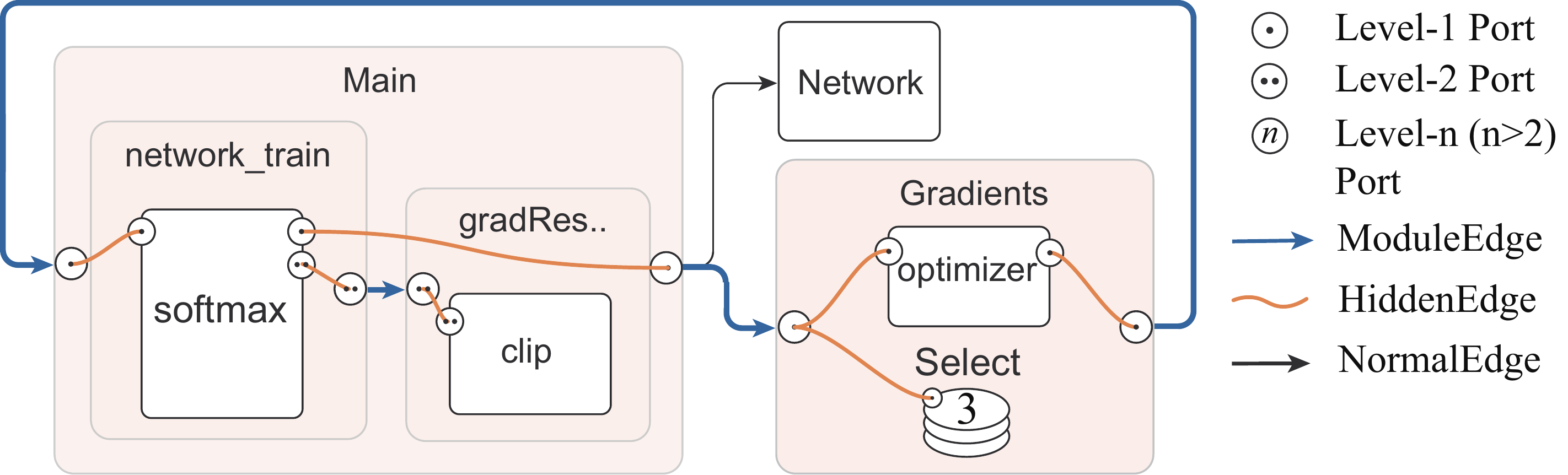}
  \vspace{-.6cm}
  \caption{Our module-based edge pruning scheme. This example graph contains two level-1 \code{ModuleNodes} and two level-2 \code{ModuleNodes}.}
  \label{fig:port-design}
\end{figure}

Each \code{ModuleNode} has an explicit attachment point called ``port" on its left (right) border, which assembles the input (output) edges inside the module. To reduce visual clutter, these edges are hidden (R1.2). Correspondingly, the other end of the hidden edges which is on a \code{NonModuleNode} is also represented as a port. The port of a \code{NonModuleNode} shares the same visual encoding with that of a \code{ModuleNode}, but is smaller. As illustrated in \figurename~\ref{fig:port-design}, two top-level nodes named \code{Main} and \code{Gradients} are recognized as \code{ModuleNodes} whose module level equals 1. Their ports are represented by glyphs like \inlinegraphics{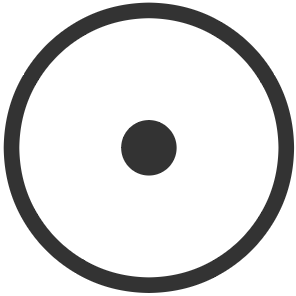}. The node inside \code{Main} named \code{network\_train} is also a \code{ModuleNode} with a level-2 port: \inlinegraphics{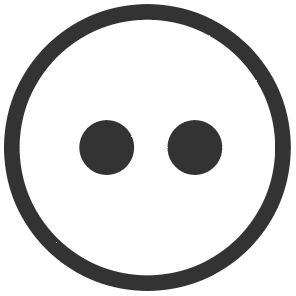}. The level-$n$ $(n\textgreater 2)$ ports are like \inlinegraphics{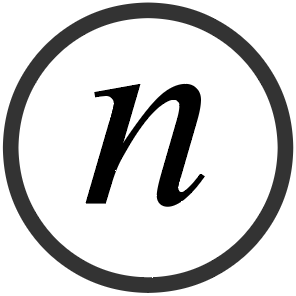}. The child node \code{softmax} of the \code{network\_train}, which is a \code{NonModuleNode}, also contains ports whose level are determined by their correspondent parent \code{ModuleNode}. The two output ports of \code{softmax} which connect to a level-1 module and a level-2 module respectively, are juxtaposed vertically on the right border.

Edges can be categorized into three types (see \figurename~\ref{fig:port-design}), by considering whether they share the same parent and whether their source node or target node are in a \code{ModuleNode} (R3.1). 
\begin{enumerate}
 \item A \code{ModuleEdge} starts on the output port of a \code{ModuleNode} and ends on the input port of another \code{ModuleNode}.
 \item A \code{HiddenEdge} connects the port of the \code{ModuleNode} and the border of its inner \code{NonModuleNode}. A \code{HiddenEdge} is represented as a Bézier curve and is invisible by default. When the port of a \code{NonModuleNode} is hovered, all its connected hidden edges are shown.
 \item The other edges are \code{NormalEdges}, whose two ends are on the node's border.
\end{enumerate}
\vspace{-.1cm}




\subsubsection{Isomorphic Subgraph Stacking}\label{subsubsec:isomorphic-subgraph-stacking}
A large amount of identical or similar structures duplicate in many computational graphs. For example, as a typical embedding model for NLP, BERT~\cite{devlin2019bert} stacks multiple layers of attention, and its computational graph becomes dominated by repeated isomorphic subgraphs of these layers. Thus, the perception and exploration of the graph become challenging. Besides, when the user attempts to expand a \code{MetaNode} containing repeated isomorphic subgraphs, it is too large to lay out interactively, which is not appropriate for visualization purposes. 

To handle isomorphic substructures, coarsening techniques are needed. TensorBoard extracts specific nodes to emphasize critical patterns at the cost of integrity and continuity. This condition results in a possible loss of the graph structure information.

An efficient strategy is required to recognize the regularity of the isomorphic subgraphs (R2.1). Retaining as much topology information (R2.2) as possible after merging or extracting the repetitive elements is also necessary. Our solution is to search isomorphic subgraphs from the entire computational graph and stack the subgraphs node by node, significantly reducing repetitive elements without sacrificing the intuitiveness of topological structures.

\begin{figure}[H]
  \vspace{-.1cm}
  \centering
  \includegraphics[width=1.0 \linewidth]{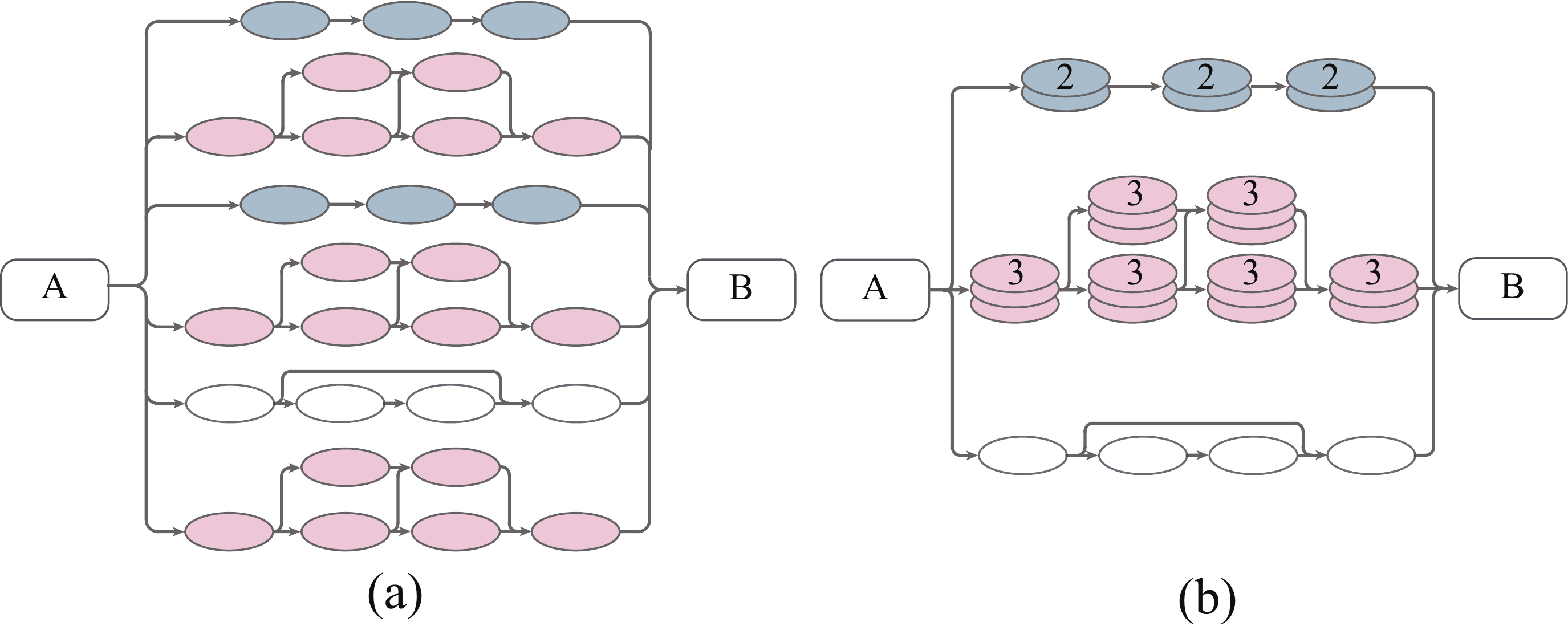}
  \vspace{-.7cm}
  \caption{The illustration of isomorphic subgraphs stacking: two groups of isomorphic subgraphs colored in blue and pink respectively in (a) are stacked into two piled subgraphs in (b), while the uncoloured subgraph remains unchanged. Counts of the stacked nodes are marked on the piles.}
  \label{fig:isomorphic-subgraph}
  \vspace{-.1cm}
\end{figure}

We first introduce how detected isomorphic subgraphs are stacked. In \figurename~\ref{fig:isomorphic-subgraph}a, subgraphs between node $A$ and $B$ are colored by isomorphic categories. \figurename~\ref{fig:isomorphic-subgraph}b shows the result of our algorithm. Two subgraphs in purple and three subgraphs in yellow are stacked into ``piles". The number on the ``pile" node denotes the repeat times of the subgraph it belongs to. The other subgraphs remain unchanged.

\begin{figure}[H]
  \vspace{-.1cm}
  \centering
  \includegraphics[width=1.0 \linewidth]{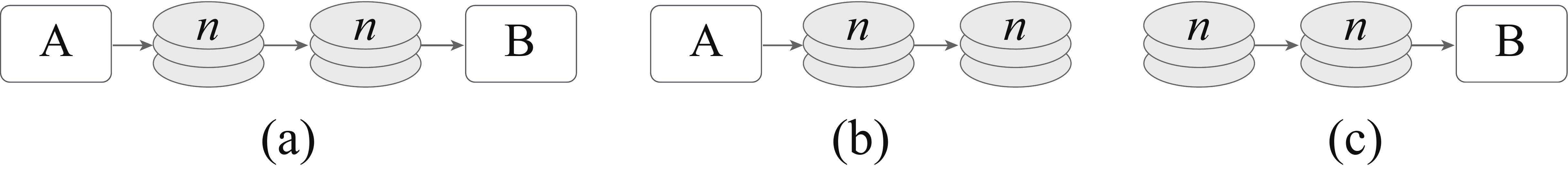}
  \vspace{-.6cm}
  \caption{The illustration of three types of isomorphic subgraphs.}
  \label{fig:isomorphic-subgraphs-categories}
  \vspace{-.1cm}
\end{figure}

With the goal of efficiently traversing isomorphic subgraphs, we classify isomorphic subgraphs as three categories: (1) subgraphs between the same source node and target node (\figurename~\ref{fig:isomorphic-subgraph}b and \figurename~\ref{fig:isomorphic-subgraphs-categories}a); (2) subgraphs that are connected with the same source node and no target nodes (\figurename~\ref{fig:isomorphic-subgraphs-categories}b); (3) subgraphs that are connected with the same target nodes and no source nodes (\figurename~\ref{fig:isomorphic-subgraphs-categories}c). We first scan all nodes to detect the isomorphic subgraphs that are connected with a source node (the first and the second category), and then detect subgraphs that are connected with no source nodes(the third category). After several rounds of traversals, a clear majority of isomorphic subgraphs in the graph are reached. The pseudocode of the detailed algorithm is displayed in Appendix C. The time complexity of the isomorphic subgraph stacking algorithm is $O(|N|^2 + |N|\cdot|E|)$. 

\begin{figure*}[h]
  \centering
  \includegraphics[width=\linewidth]{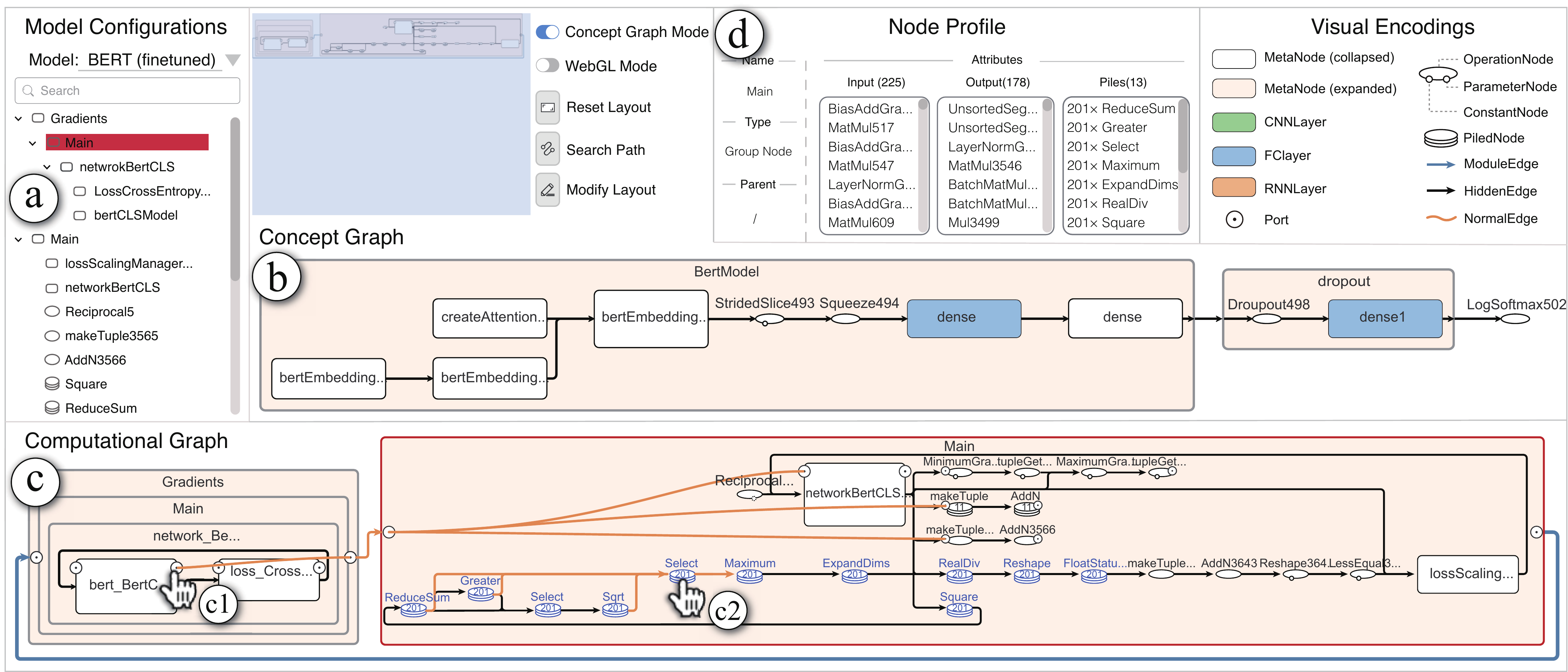}
  \caption{Our visual interface with the example of BERT~\cite{devlin2019bert}: (a) A model configuration view supports selecting a model and shows the hierarchical structure of its computational graph. (b) The concept graph of BERT shows its forward propagation graph structure with the cycles removed. The FC layers like ``dense'' are simplified. (c) The main view illustrates visual simplifications of the computational graph. Here, a BERT model is employed. The node and edge numbers are reduced from 4052 and 6353 to 55 and 74, respectively. By applying port constraints, edges are divided into bundled segments and curve segments. Two of the interactions are illustrated here. Hovering on a port highlights its connected edges in orange (c1); The isomorphic subgraphs are highlighted in blue when one of the piled nodes is hovered (c2). (d) The node profile view lists topological structures and configuration information of the selected node \code{Main}.}
  \label{fig:system-interface}
\end{figure*}

To effectively recognize and stack isomorphic subgraphs, we design a hash-based algorithm to represent each detected subgraph. The hash value of a subgraph $g=(N, E)$ is calculated as:
$h_g(g) = [ \sum\limits_{i=1}^{|N|} h_n(n_i) + \sum\limits_{i=1}^{|E|} h_e(e_i) ] \bmod P$.
The hash functions $h_n$ and $h_e$ are based on the DJB algorithm~\cite{estebanez2014performance}, which excels at string hashing. Here, P is set at 10000019 to avoid data overflow. For each node $n_i \in N$, we establish a string set $S_{n_i}$ composed of $n_i$'s type, neighbors type, parent id, indegree, outdegree, and \#(auxiliary nodes). The hash value of $n_i$ is defined as:
$h_n(n_i) = \sum\limits_{j \in S_{n_i}} DJB(n_{ij}) \bmod P$.
Each edge $e_i \in E$ is encoded as a string $s_{e_i}= $``\code{source\_type} $\rightarrow$ \code{target\_type}", where \code{source\_type} represents the type of the source node and \code{target\_type} represents the type of the target node. The hash value of $e_i$ is defined as:
$h_e(e_i) = DJB(s_{e_i}) \bmod P$.


\subsection{User Exploration}
The computational graph visualization tool is shown in \figurename~\ref{fig:system-interface}. The model configuration view (a) allows the user to select a model and displays the hierarchical structure of the model's computational graph. By turning on the Concept Graph Mode, the concept graph (b) shows the forward propagation graph. Otherwise, the simplified computational graph layout (c) is given. The user can freely expand, collapse, and ungroup a \code{MetaNode}. Dragging an \code{OperationNode} modifies the layout automatically. A stable transition animation is supported when the layout changes. The related bundled edges are highlighted (c1), and all the connected hidden edges are shown (c2) by hovering a port. 

In some extreme scenarios, the user may feel confused due to the complexity of connections. To avoid such situations, the user is allowed to check certain connections separately, as follows: 1) the hidden edges can be highlighted by hovering the corresponding port; 2) the connected edges can be highlighted by selecting one or two nodes as endpoints in the graph. In addition, nodes can be searched in (a) by inputting a name, such as \code{clip\_gradients\_ClipGradients}. Then, the node profile (d) is shown, and the connected edges are highlighted.

\subsection{Implementation} ~\label{subsection:implementation}
We implement a web-based visual interface (as shown in \figurename~\ref{fig:system-interface}), powered by Django and ELK. ELK provides a layer-based layout algorithm that is suited for node--link diagrams with an inherent direction and ports. A computational graph is generated by recursively calculating the port--constraint orthogonal layout for each expanded \code{MetaNode}'s subgraph (visible graph). We implement a stable transition animation when the layout changes on the basis of the ``interactive mode" provided by ELK. In particular, we propose to insert frames nonlinearly between two frames, leading to a smoother result.

We use SVG and D3.js~\cite{BostockOH11} for graph visualization and employ  PixiJS\footnote{https://www.pixijs.com} to implement a WebGL~\cite{parisi2012webgl} version. PixiJS is applied because of its rich APIs to render interactive graphics. 

\begin{figure*}[h]
    \centering
    \includegraphics[width=1.\linewidth]{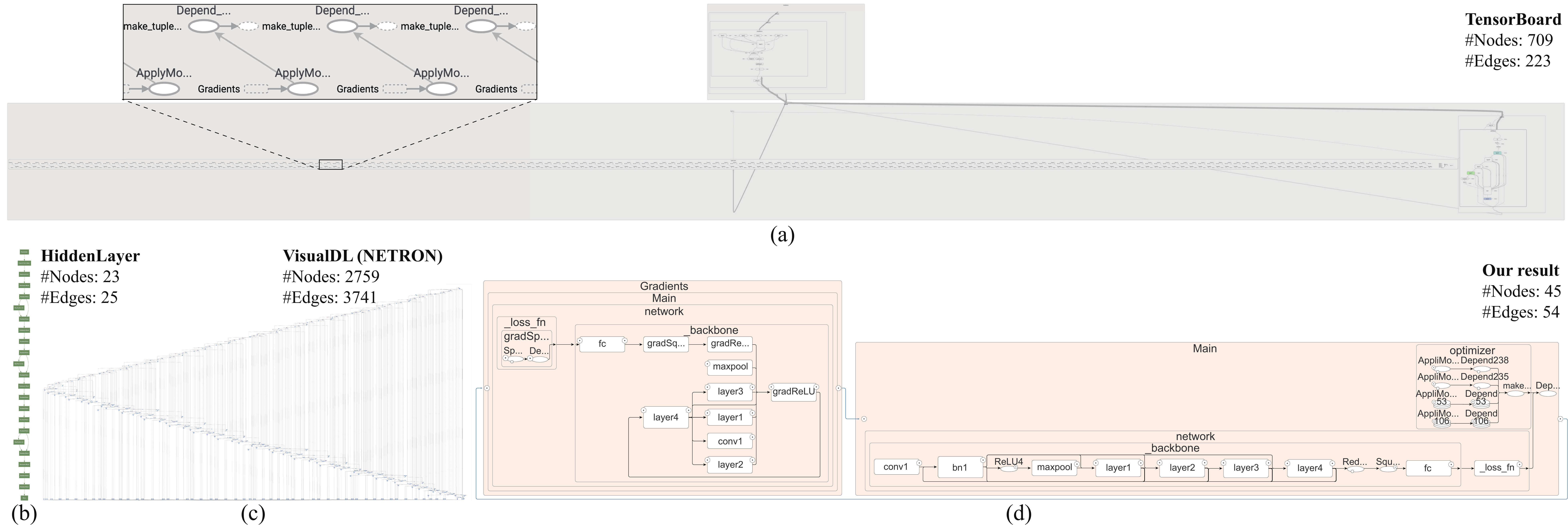}
    \vspace{-.4cm}
    \caption{The computational graphs of ResNet-50 yielded by different tools: (a) Tensorboard, (b) HiddenLayer, (c) VisualDL, and (d) ours.}
    \label{fig:evaluation-resnet}
    \vspace{-.3cm}
  \end{figure*}


\section{Evaluation}\label{sec:evaluation}
To quantify how our system can simplify the computational graphs of DNNs and demonstrate its performance for recognizing and diagnosing DNN models, we present a set of experiments including a quantitative analysis, two usage scenarios, and a user study containing four typical tasks. We summarize the feedback from the domain experts.
\subsection{Quantitative Analysis}
\textbf{Setup.} The experiments are performed on a Windows PC with an Intel Core i7-9700 CPU @ 3.00GHz, 16GB RAM, and Geforce GTX 1660 Ti. The system interface is displayed on a 27-inch screen (2,560 $\times$ 1,440), using the Google Chrome browser.

\textbf{Experiments.} We compare our approach (the online system\footnote{https://mindinsight1.natapp1.cc/} is accessible) with existing methods on different computational graph data. Among the existing methods, Net2Vis, draw\_convnet, and convnet\_drawer only supports visualizing the computational graph of CNNs as schematic diagrams. They are not applicable for other DL models. Their results provide an overview of graphs. Interactions for exploring the internal details of layers and the holistic data flow are not supported. HiddenLayer supports more types of DL models, but only provides a static overview, where interactions like expanding or collapsing metanodes are not supported. TensorBoard and VisualDL (powered by NETRON) allow users to interactively manipulate with heterogeneous models. Consequently, to evaluate the results of various networks both on high-level architectures and the detailed topologies, we compare our approach with TensorBoard, HiddenLayer, and VisualDL (NETRON). Two mainstream models generated from MindSpore are used as test datasets: ResNet-50~\cite{he2016resnet} (1,120 operators and 1,629 connections), and BERT (5,698 operators and 7,340 connections).

\figurename~\ref{fig:evaluation-resnet} displays the computational graphs of ResNet-50 created by the four tools. In the TensorBoard output(\figurename~\ref{fig:evaluation-resnet}a), severe visual clutter is caused by a high number of repetitive subgraphs. The enlarged part is shown on the upper left corner. The situation with VisualDL (NETRON) is much worse (\figurename~\ref{fig:evaluation-resnet}c), where the hierarchy information is not displayed correctly. The result of HiddenLayer (\figurename~\ref{fig:evaluation-resnet}b) reveals the model's high-level design but no layer-level details. The generated static image looks long and narrow, which is hard for users to keep the mental map when observing the detailed connections by zooming and panning. In the computational graph using our algorithm (\figurename~\ref{fig:evaluation-resnet}d), repetitive subgraphs are stacked into four piles on the upper right corner, and the topologies are well maintained. Similarly, the results of BERT are demonstrated in Figure~\ref{fig:intro-bert}. The VisualDL and HiddenLayer both fail to yield the computational graphs because of running out of memory.

\begin{figure}[H]
    \centering
    \includegraphics[width=1.\linewidth]{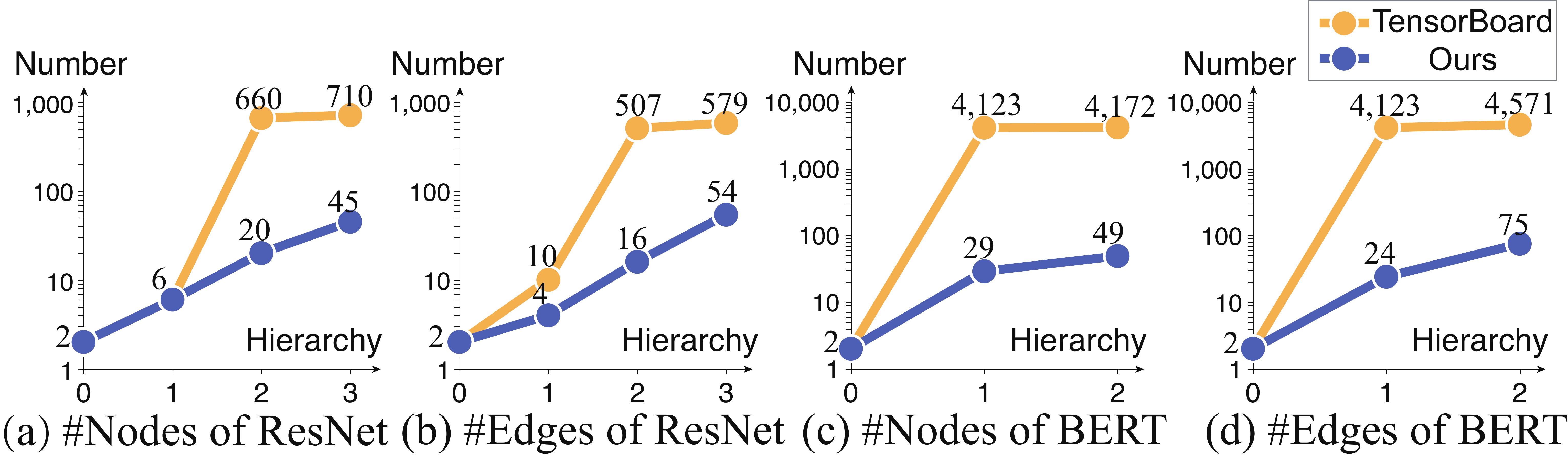}
    \caption{Numbers of nodes and edges after expanding each hierarchy with TensorBoard and our tool. Ours reduces 60\% elements on average.}
    \label{fig:evaluation-number}
\end{figure}

As previously mentioned, HiddenLayer does not support expanding hierarchies, and VisualDL (NETRON) fails to display the hierarchies correctly. Consequently, we compare TensorBoard and our method by counting nodes and edges after expanding each hierarchy. As indicated by \figurename~\ref{fig:evaluation-number}a, the total number of nodes in the third hierarchy is dramatically reduced from 710 to 45. The module-based edge-pruning scheme leads to a decrease in the edge number from 579 to 54 (\figurename~\ref{fig:evaluation-number}b). Users can expand more nodes and check details of more hierarchies on the basis of our results. \figurename~\ref{fig:evaluation-number}c and d show the result on BERT. When two hierarchies are available, the computational graph from TensorBoard consists of over 4,000 elements, but our result contains less than 100 elements. 


\begin{figure}[h]
    \centering
    \includegraphics[width=.8\linewidth]{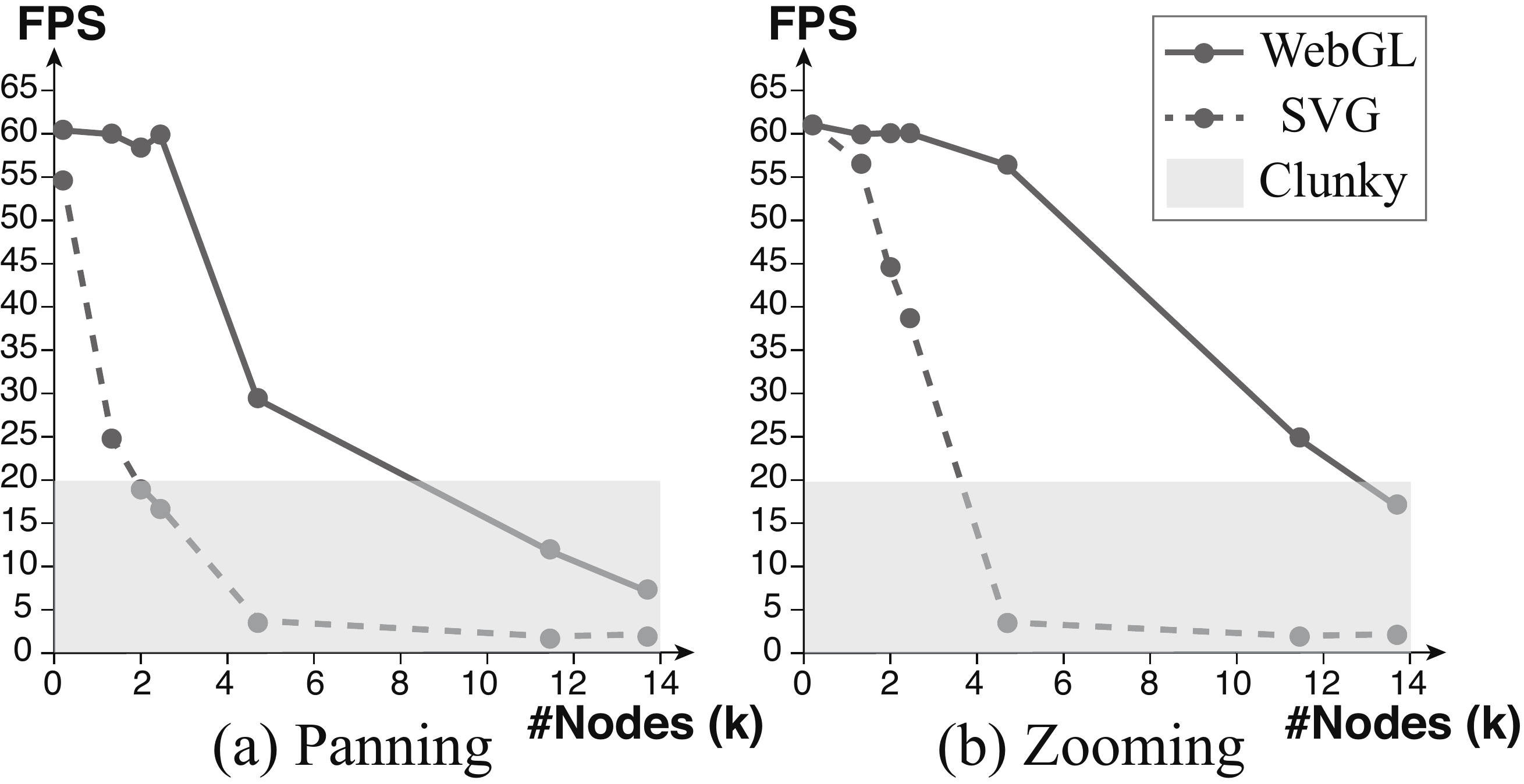}
    \caption{A synthetic graph with 884 links and 138 ports is used for the test initially. We add nodes into the graph until 14,000 to observe the FPS of panning and zooming interactions rendering with SVG and WebGL, respectively.}
    \label{fig:evaluation-fps}
\end{figure}

Moreover, we compare the interaction performance of the WebGL version and SVG version in \figurename~\ref{fig:evaluation-fps}. A synthetic graph for the test contains 884 links and 138 ports. The results show that the WebGL version supports smooth interactions ($\geq$ 20FPS) for panning up to 5 thousand nodes and zooming with up to 11 thousand nodes.

\subsection{Usage Scenarios}

We describe two usage scenarios using different DNN datasets. The first case focuses on exploring the architecture of a complex model, and the second one illustrates how our approach assists model diagnosis with the simplified computational graph.

\begin{figure*}[h]
    \centering
    \includegraphics[width=1.\linewidth]{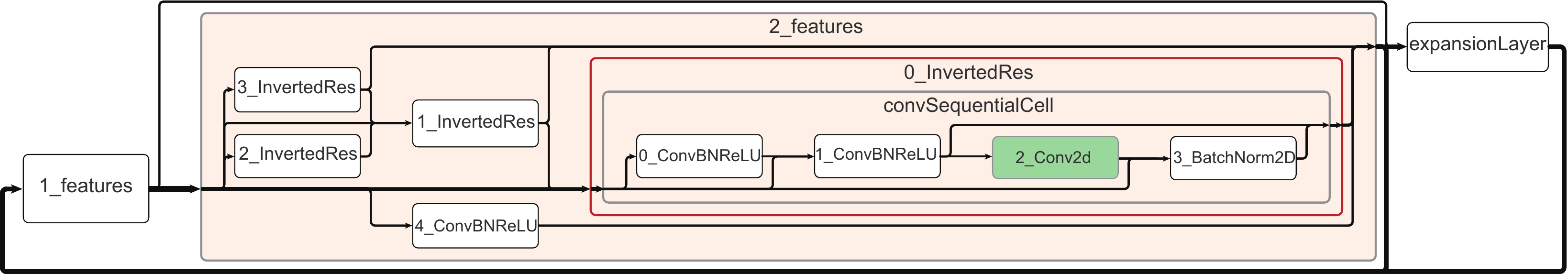}
    \caption{The concept graph of SSD300. The expansion layer and inverted residual structures indicate that the backbone network of the SSD300 is MobileNetV2.}
    \label{fig:ssd300-cgm}
\end{figure*}

\subsubsection{Scenario 1: Learning model structures}
SSD (Single Shot MultiBox Detector)~\cite{DBLP:conf/eccv/LiuAESRFB16} is a high-performance object detection model, which is constructed based on a backbone network (e.g., VGG16~\cite{DBLP:journals/corr/SimonyanZ14a}, MobileNetV2~\cite{sandler2018mobilenetv2}, and ResNet~\cite{he2016resnet}). A male developer uses SSD300 for the first time. To learn the architecture of object detection intuitively and efficiently find out which backbone network is used, he utilizes our system to visualize the computational graph of SSD300 ($\textgreater 5k$ nodes, $\textgreater 7k$ edges).

\begin{figure}[H]
    \centering
    \includegraphics[width=1.\linewidth]{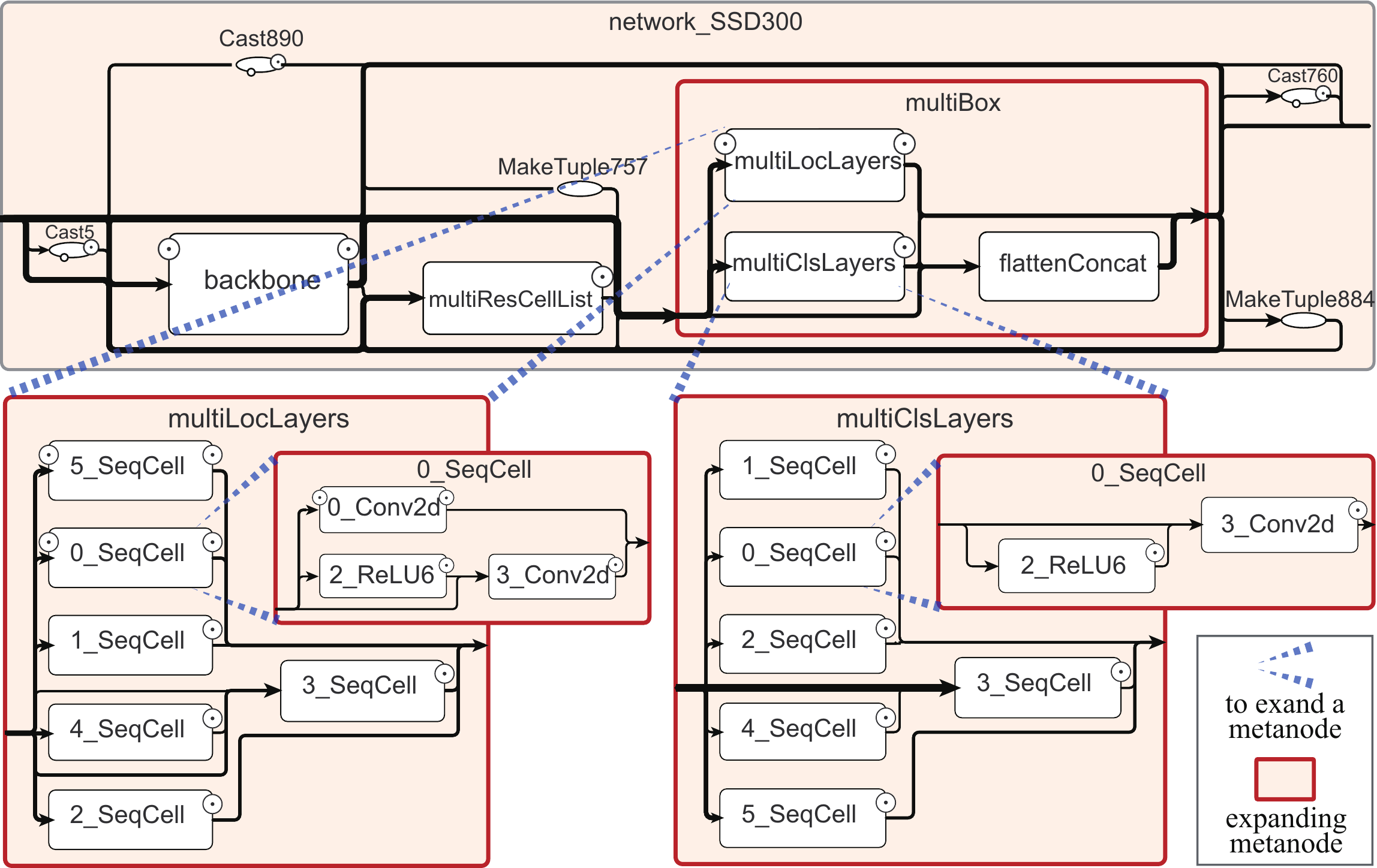}
    \caption{The object detection layers of SSD300. In the metanode \code{multiBox}, both \code{multiLocLayers} and \code{multiClsLayers} contain six sequential cells. However, their sequential cells have different structures to perform different tasks.}
    \label{fig:ssd300-network}
\end{figure}

The user wants to explore object detection structures. The original computational graph is composed of two top-level \code{MetaNodes}, namely, \code{Gradients} and \code{Main}, which are automatically defined by MindSpore. \code{Main} denotes the user-defined network structures including optimizers, and \code{Gradient} represents the gradient calculation of the parameters from \code{Main}. To overserve the forward propagation network, xhe expands the node \code{Main} recursively until the node \code{network\_SSD300} (see \figurename~\ref{fig:ssd300-network}). The node \code{backbone}, \code{multiResidualCellList} and \code{multiBox} are contained. After expanding \code{multiBox}, \code{multiLocLayers} (used to output coordinates) and \code{multiLClsLayers} (used to predict classes) appear. Each layer contains six sequential cells. To compare them, he expands \code{0\_SeqCell} in each layer respectively and the differences are shown.

To identify the used backbone network, the user observes the computational graph in the CGM (see \figurename~\ref{fig:ssd300-cgm}). The generated computational graph shows a forward propagation graph structure starting with two feature layers and ending with an expansion layer. Because MobilenetV2 contains an expansion layer, they infer that the backbone network is MobilenetV2. For more evidence, he expands the node \code{2\_Features}. A series of nodes named \code{X\_InvertedRes} are contained. \code{Inverted Residual} is proposed by MobilenetV2 as a representative module. By expanding the \code{0\_InvertedRes} in \code{2\_Features}, the residual structure is shown. After checking the channels of the convolution layers, he finds that the intermediate layers have a higher number of channels. Consequently, he is convinced that the expansion layer and inverted residual layers are used. As a result, the backbone network is the MobileNetV2.

\subsubsection{Scenario 2: Optimizing a DNN} \label{subsubsection:Scenario2}


\vspace{-.3cm}
\begin{table}[h]
    \centering
    \caption{The time consumption (before optimization).}
    \vspace{-.3cm}
    \begin{tabular}{lc}
        \toprule
        op\_name                           & cuda\_cost\_time($\mu$s) \\ \midrule
        \rowcolor[HTML]{a1afbf} backbone/Conv2/Conv2D-op211          & 169.087                               \\
        backbone/Conv1/Conv2D-op207          & 135.391                               \\
        backbone/FC3-Dense/FusedMatMulBiasAdd-op298 & 56.384                               \\ \bottomrule
    \end{tabular}
    \label{table:operator-rank}
\end{table}

Another male developer is analyzing the operator time-consumption rankings during the training process of a LeNet network. As indicated in Table~\ref{table:operator-rank}, the operator named \code{backbone/Conv2/Conv2D-op211} consumes the most time on average. However, because the operator names generated by the profile are different from the ones defined by the user, it is hard to directly locate the variable declaration in the user's code, in which two conv2D layers are declared. 

\begin{figure}[h]
    \centering
    \includegraphics[width=1.\linewidth, trim=0 0 0 90, clip]{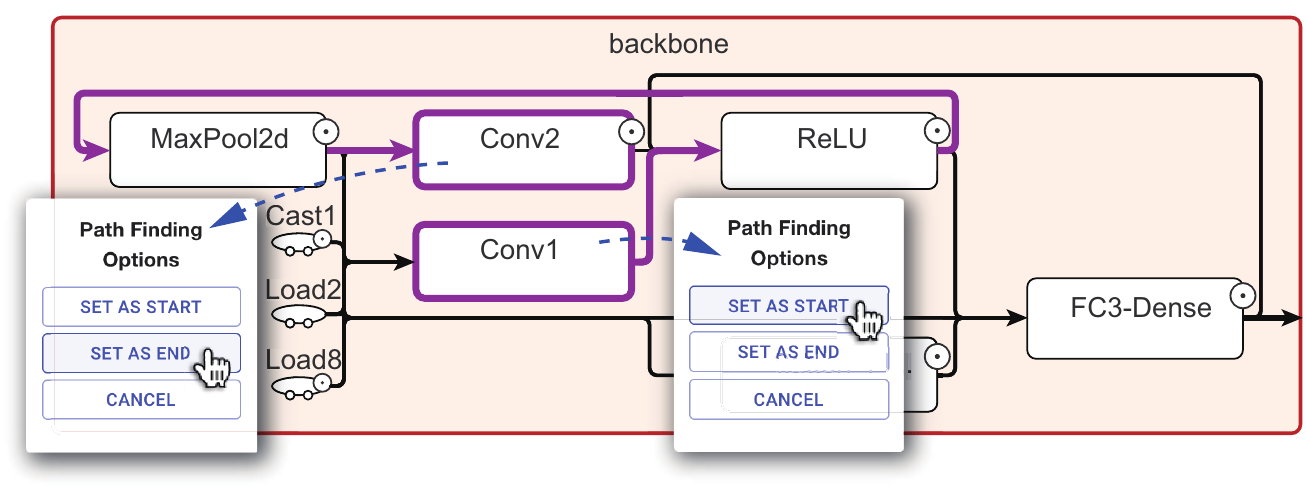}
    \vspace{-.9cm}
    \caption{In the path-finding mode, the user sets \code{Conv1} as start node and \code{Conv2} as end node. If at least one path from the start node to the end node exists, the path will be highlighted in bold and colored in purple.}
    \label{fig:lenet-path}
\end{figure}

\begin{figure}[h]
    \centering
    \includegraphics[width=1.\linewidth]{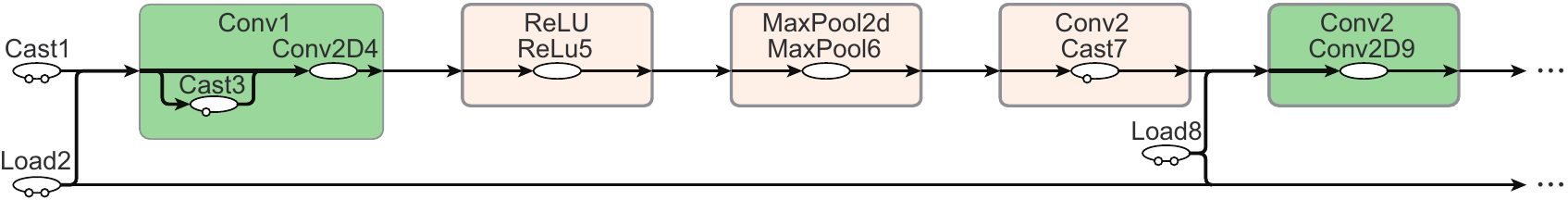}
    \caption{The concept graph of the network to be optimized.}
    \label{fig:lenet-cgm}
\end{figure}

On the basis of this, he uses our tool to scrutinize the operator node \code{Conv1} and \code{Conv2} in the computational graph. Suppose that \code{Conv1} is executed before \code{Conv2}, a directed path from \code{Conv1} to \code{Conv2} exists. To verify the hypothesis, he turns on the path-finding mode, which highlights the possible paths between a given start node and end node. Then he sets \code{Conv1} as the start node and \code{Conv2} as the end node. The path between them appears, which means that \code{Conv1} precedes \code{Conv2}. \figurename~\ref{fig:lenet-path} illustrates the process. Alternatively, using the concept graph mode (as indicated in \figurename~\ref{fig:lenet-cgm}) can make the order between \code{Conv1} and \code{Conv2} clearer.

\vspace{-.1cm}
\begin{table}[h]
    \centering
    \caption{The time consumption (after optimization).}
    \vspace{-.3cm}
    \begin{tabular}{lc}
        \toprule
        op\_name                           & cuda\_cost\_time($\mu$s) \\ \midrule
        backbone/Conv1/Conv2D-op207          & 134.526                              \\
        backbone/FC3-Dense/FusedMatMulBiasAdd-op298 & 54.304                               \\ 
        \rowcolor[HTML]{a1afbf} backbone/Conv2/Conv2D-op211          & 109.631                               \\
        \bottomrule
    \end{tabular}
    \label{table:operator-rank2}
\end{table}
\vspace{-.1cm}

With the help of the computational graph, the user successfully locates the operator. He finds that the numerical precision of \code{Conv2} is FP32 while that of \code{Conv1} is FP16. He infers that a high precision leads to the bottleneck of performance. As a result, he sets \code{Conv2} from FP32 to FP16 and retrains the network. As indicated in Table~\ref{table:operator-rank2}, the time consumption of \code{Conv2} decreases dramatically from 169.087$\mu$s to 109.631$\mu$s with others' time nearly unchanged.

\subsection{User Study}

The goal of our study is to compare our tool with TensorBoard by investigating whether users can intuitively identify the structures and features of complex models, as well as to efficiently diagnose the code of the model. To make it convenient for remote testing because of COVID-19, we adopt the between-subjects design with randomizing the sequence of tasks. Each participant is required to use one of the two systems to visualize a series of computational graphs generated by MindSpore.

\subsubsection{Procedure and Tasks}
Before the study, we introduce the characteristics of the system to the participants. All the interactions are illustrated in detail to instruct them to explore the example graph dataset using the system. The background knowledge needed for finishing the tasks are attached. Participants are required to finish the tasks online remotely using a Chrome browser on a screen with a resolution of 1,920 $\times$ 1,080. The computational graph datasets are preloaded before recording the results. Each participant's answers to the tasks, the time cost, ratings of the questionnaire, and personal feedbacks are recorded. 

Twenty-four volunteers (eighteen males and six females) with an average of 2.458 ($SD=1.450$) years of experience in DL are recruited. One-third of them are DL engineers, and the others are graduate students who study DL. The participants are required to perform the following tasks.

After discussing with domain experts and senior developers, we design two types of tasks: cognitive tasks and diagnostic tasks.

\textbf{C}ognitive \textbf{T}asks require participants to understand the architecture of DNNs by exploring the computational graph, and include two tasks:
\begin{itemize}
    \item \textbf{CT1 (R1, R2):} Identify the genealogy of given DNNs. The participant is required to explore the computational graphs of two anonymous networks (ResNet-50 and Inceptionv3) and recognize their network genealogies. Four options are provided, including Bert, ResNet~\cite{he2016resnet}, Inception~\cite{DBLP:conf/cvpr/SzegedyVISW16}, and LSTM~\cite{DBLP:conf/nips/ShiCWYWW15}. 
    \item \textbf{CT2 (R3):} Estimate the depth of given DNNs. The typical ResNet networks include 18-layers, 34-layers, 50-layers, 101-layers, and 152-layers. Each of them is characterized with different convolution structures. The participant is required to explore the computational graph of an anonymous ResNet (ResNet-34) and answer the count of layers (34-layers).
\end{itemize}

\textbf{D}iagnostic \textbf{T}asks require participants to analyze the computational graph according to the code of the DNN model, and include two tasks:
\begin{itemize}
    \item \textbf{DT1 (R1, R3):} Check whether the code of DNN training model matches the computational graph. Firstly, a piece of code of model definition is given, where the function \code{stop\_gradient} is used to prevent computing gradient of some layers during back propagation. The participant needs to understand which layers are stopped according to the given code. Then they need to observe the corresponding structures in the computational graph to check whether the correct layers are stopped. If not, they need to identify where the \code{stop\_gradient} is used in the code of the given computational graph.
    \item \textbf{DT2 (R2):} Locate the operator in the profile. This task is designed on the basis of Section~\ref{subsubsection:Scenario2}. The participants are first given the ranking of operators and part of the code of model definition to learn about the model architecture. Then they need to interact with the computational graph to observe the edges between \code{Conv1} and \code{Conv2} to give their ranking. 
\end{itemize}

\begin{table}[H]
    \vspace{-.3cm}
    \centering
    \caption{The sequence of tasks for different groups.}
    \vspace{-.3cm}
    \begin{tabular}{@{}c|cccc|cccc@{}}
    \toprule
    Condition & \multicolumn{4}{c|}{Our System} & \multicolumn{4}{c}{TensorBoard}  \\ 
    \midrule
    Group     & G1 & G2  & G3  & G4  & G5 & G6  & G7 & G8  \\ 
    \midrule
    Task 1    &\textbf{CT1} & \textbf{CT2} & \textbf{DT1} & \textbf{DT2} &\textbf{CT1} & \textbf{CT2} & \textbf{DT1} & \textbf{DT2} \\
    Task 2    & \textbf{CT2} & \textbf{DT1} & \textbf{DT2} &\textbf{CT1} & \textbf{CT2} & \textbf{DT1} & \textbf{DT2} &\textbf{CT1} \\
    Task 3    & \textbf{DT1} & \textbf{DT2} &\textbf{CT1} & \textbf{CT2} & \textbf{DT1} & \textbf{DT2} &\textbf{CT1} & \textbf{CT2} \\
    Task 4    & \textbf{DT2} &\textbf{CT1} & \textbf{CT2} & \textbf{DT1} & \textbf{DT2} &\textbf{CT1} & \textbf{CT2} & \textbf{DT1} \\ 
    \bottomrule
    \end{tabular}
    \label{table:conditions}
    \vspace{-.1cm}
\end{table}

Required by the between-subjects design, each subject only uses one of our system and TensorBoard to finish all the tasks. Our participants are firstly segmented into two parts, using our system or TensorBoard respectively. Each participant is only exposed to one of the systems, and does not know which system they are using. To avoid the sequential effects of the tasks and the variance of the participants, we adopt a 4 $\times$ 4 Latin Square to allocate the tasks. Following the rule of Latin Square, the participants of each part are further divided into four groups, of which each group takes a column of task sequence in the Latin Square. Finally, all the participants are divided into eight groups. The detailed tasks for eight groups are indicated in Table~\ref{table:conditions}. Each group contains three participants.

\begin{table}[H]
    \vspace{-.2cm}
    \centering
    \caption{Questionnaire}
    \vspace{-.3cm}
    \begin{tabular}{@{}l|l@{}}
    \toprule
    Q1 & The system is easy to learn. \\
    Q2 & The system is easy to use. \\
    Q3 & The layout is balanced between overview and detail. \\
    Q4 & The layout helps me effectively understand the model. \\
    Q5 & The interaction provided by the system is useful. \\
    Q6 & The interaction provided by the system is smooth. \\
    Q7 & The visual design is intuitive. \\
    Q8 & The visual design is good-looking. \\
    Q9 & I am confident in the solutions for the tasks. \\
    Q10 & I am willing to use the system for future development. \\ \bottomrule
    \end{tabular}
    \label{table:questionnaire}
    \vspace{-.1cm}
\end{table}

To collect subjective feedback of our tasks to understand users' preferences and directions for improvements, we design a questionnaire (as listed in Table~\ref{table:questionnaire}) containing ten questions in seven-point Likert scales. Finally, the personal feedback on different tools is collected from the participants. 

\subsubsection{Results}
We analyze the results of the accuracy, time consumption, and user feedback.

\begin{figure}[H]
    \centering
    \includegraphics[width=.9\linewidth]{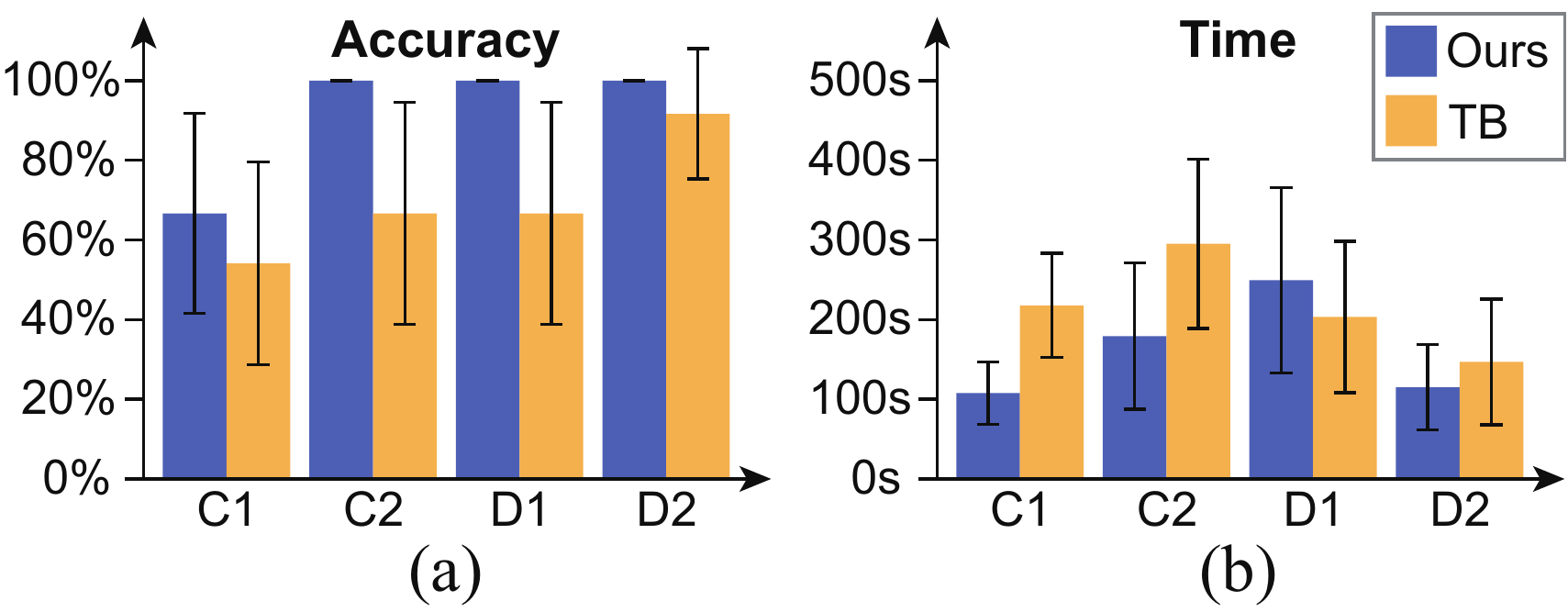}
    \vspace{-.3cm}
    \caption{(a) The average accuracy and its 95\% confidence intervals. (b) The time cost and its 95\% confidence intervals}
    \label{fig:user-study-result-cases}
\end{figure}

\textbf{Accuracy.} 
We use the paired samples t-test to examine the results. The results of the average accuracy are presented in \figurename~\ref{fig:user-study-result-cases}a. It can be seen from the first row in Table~\ref{table:tasks-ttest} that the accuracy of \textbf{CT2} and \textbf{DT1} shows a significant difference between ours and TensorBoard. It is apparent that the accuracy of our tool on these two tasks both outperforms TensorBoard. Moreover, the average accuracy of our tool on \textbf{CT1} and \textbf{DT2} is higher than that of TensorBoard, though no significant difference exists. Benefited from the simplified representations and utility modes of our tool, users can recognize the structures of the computational graphs more intuitively.

\begin{table}[H]
    \vspace{-.2cm}
    \centering
    \caption{The t-test results of the tasks performed on different tools. The P-value that $\textless 0.05$ means a significant difference, which is highlighted in bold.}
    \vspace{-.3cm}
    \begin{tabular}{c|cccc}
        \toprule
        P-value & \textbf{CT1} & \textbf{CT2} & \textbf{DT1} & \textbf{DT2} \\ \midrule
        Accuracy  & 0.501   & \textbf{0.028} & \textbf{0.028} & 0.328 \\
        Time  & \textbf{0.010}  & 0.120 & 0.556 & 0.518  \\ \bottomrule
        \end{tabular}
    \label{table:tasks-ttest}
    \vspace{-.2cm}
\end{table}

\textbf{Time.} 
The average costing time of \textbf{CT1} (the second row of Table~\ref{table:tasks-ttest}) verifies the advantages of our system over Tensorboard. On the other tasks except \textbf{DT1}, our tool costs notably less time than that of TensorBoard. Thanks to the faster renderer and fewer rendering elements, our tool saves a great deal of time on transition and other interactions.

\begin{figure}[H]
    \centering
    \includegraphics[width=.9\linewidth]{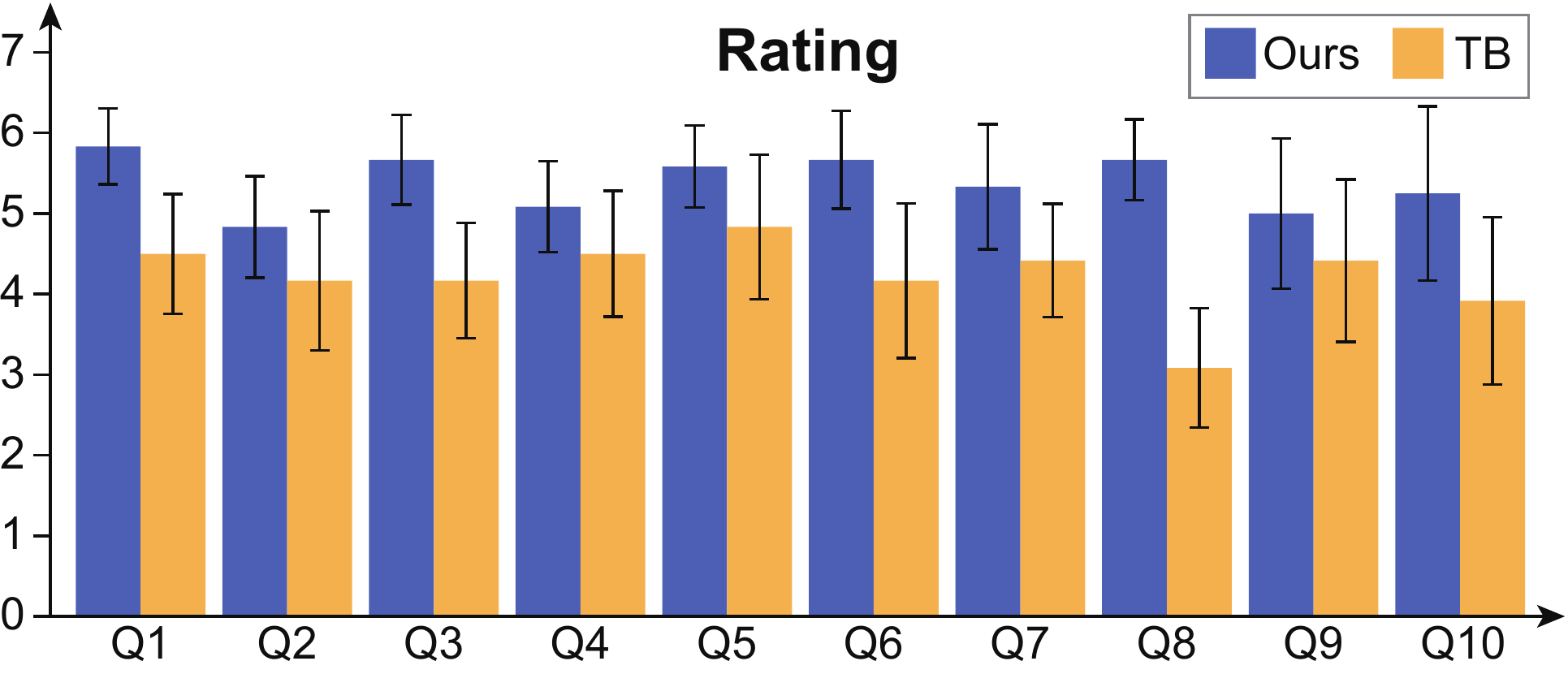}
    \vspace{-.3cm}
    \caption{The average ratings of the questionnaire and 95\% confidence intervals.}
    \label{fig:user-study-result-questionnaire}
\end{figure}

\textbf{Questionnaire.} 
\figurename~\ref{fig:user-study-result-questionnaire} shows the ratings of the questionnaires. In summary, the users prefer our system ($mean=5.391, SD=0.323$) to TensorBoard ($mean=4.217, SD=0.448$). It indicates that our system is easier to use, the visualization is more balanced and effective, and the design is more intuitive. 

\subsection{Expert Feedback}

We collect feedback from five domain experts from MindSpore and MindInsight communities. First, we introduce the essential functions of our prototype system by demonstrating the usage scenarios to show the usage of our system. Thereafter, we invite them to follow the workflow and use our prototype system for one day. Finally, we interview them and summarize their feedback.

\textbf{The overall evaluation is quite positive}: \emph{The synthetic layout strategies achieve a balance between effectiveness and efficiency. All the elements of the entire computational graph are well organized with an orthogonal-style layout. Benefiting from the means of simplification, large-scale computational graph datasets can be loaded quickly and interacted effectively.}

\textbf{The performance of our implementation is satisfying.} On the basis of our illustration of the methods, their comments are:
\emph{
The innovative edge-pruning scheme clearly shows the relationships between the namespace and distinguishes the connective edges between multiple namespaces and the same operators.
The scene that many operators or substructures share repeated types in the same namespace is typical in complex MindSpore neural networks. Directly visualizing them tends to clutter the screen with too many redundant elements. The strategy of stacking isomorphic subgraphs solves this problem well.}

\textbf{They appreciate the interactions of our system.} Combining their usage of our tool and their domain experience, their feelings are:
\emph{Interactive ways of quickly finding target structure, including substructure search, enable developers to view large models easily. The quick response of interaction improves the exploration efficiency and user experience.}

\textbf{Some suggestions are reported}.
The layout scheme for the computational graph significantly simplifies the graph layout effect and solves the problem of edge intertwining and cluttering. However, connective relationships may be confusing after merging the same namespace edges. They recommend highlighting the connective edges by hovering and searching interactions. It is suggested that edges that may lead to semantic confusion and visual ambiguity are not removed to balance conciseness and clearness.

\section{Discussion}\label{sec:discussion}
In this section, we summarize lessons learned from our trials, and reflect on the failure cases and limitations.
\subsection{Lessons learned} 
Initially, we tried to simplify graphs on the basis of the traditional Sugiyama-style flow layout. However, it is not scalable for large datasets and is unstable for various DNNs. We noticed that the standard subway route is concise and clear. Meanwhile, in the circuit diagrams, wires are either horizontal or vertical. Benefited from this orthogonal constraint, electronic components of the circuit diagram are organized orderly. In addition, an orthogonal layout makes better usage of the rectangular screen space than other approaches. Inspired by these observations, we choose to use orthogonal-edge constraints.

The ``pile" nodes are used to encode stacked isomorphic subgraphs. This representation acquires a balance between keeping the details of the local subgraphs and global topology integrity. However, by stacking the subgraphs into a series of nodes, boundaries between different isomorphic subgraphs are vague. The connected ``pile" nodes with the same repeat number are recognized to belong to the same isomorphic subgraph. To eliminate the potential ambiguity, hovering on a ``pile" node yields the highlighting of its isomorphic subgraphs.
\subsection{Failure Cases and Limitations} 
Some failure cases are observed. The design shown in \figurename~\ref{fig:edge-design}b does not tackle all the ambiguity problems of connection relationships. For instance, a complete bipartite graph exists between $S_i (i=1,2,3,4)$ and $T_i (i=1,2)$, where each circular/straight-line segment is shared by two or more potential connection relationships. In this extreme situation, users need to hover on nodes to observe the actual connections because we cannot guarantee a connection between $S_i (i=1,2,3,4)$ and $T_i (i=1,2)$ exists. Moreover, this structure usually appears in the FC layer of DNN, encoded by the icon of the FC layer in the CGM.

Our cycle-removing algorithm can remove almost all cycles at the top level. However, it is difficult to remove all cycles iteratively. By splitting one metanode in the cycle into two metanodes, the number of the metanodes will increase exponentially if the algorithm is executed iteratively.

\section{Conclusion}\label{sec:conclusion}
This paper presents a new visual simplification approach to address the challenges in visualizing large-scale computational graphs of DNNs. Experimental results and comparisons with existing works show that our approach achieves much better structural readability and comprehensibility over prevalent solutions. We implement a prototype system, which is designed explicitly for MindSpore and is integrated into an open-source framework. In the future, we plan to test and improve the potential of our approach in modulating and optimizing DNNs.

\ifCLASSOPTIONcompsoc
  \section*{Acknowledgments}
\else
  \section*{Acknowledgment}
\fi
The authors would like to thank the reviewers for their valuable feedback. We would also like to thank the experts who we consulted/interviewed for this work. This paper is supported by the National Natural Science Foundation of China (62132017) and the Fundamental Research Funds for the Central
Universities (226-2022-00235).


\ifCLASSOPTIONcaptionsoff
  \newpage
\fi



\bibliographystyle{IEEEtran}
\bibliography{IEEEabrv}

\begin{thebibliography}{10}
\providecommand{\url}[1]{#1}
\csname url@samestyle\endcsname
\providecommand{\newblock}{\relax}
\providecommand{\bibinfo}[2]{#2}
\providecommand{\BIBentrySTDinterwordspacing}{\spaceskip=0pt\relax}
\providecommand{\BIBentryALTinterwordstretchfactor}{4}
\providecommand{\BIBentryALTinterwordspacing}{\spaceskip=\fontdimen2\font plus
\BIBentryALTinterwordstretchfactor\fontdimen3\font minus
  \fontdimen4\font\relax}
\providecommand{\BIBforeignlanguage}[2]{{%
\expandafter\ifx\csname l@#1\endcsname\relax
\typeout{** WARNING: IEEEtran.bst: No hyphenation pattern has been}%
\typeout{** loaded for the language `#1'. Using the pattern for}%
\typeout{** the default language instead.}%
\else
\language=\csname l@#1\endcsname
\fi
#2}}
\providecommand{\BIBdecl}{\relax}
\BIBdecl

\bibitem{devlin2019bert}
J.~Devlin, M.-W. Chang, K.~Lee, and K.~Toutanova, ``{BERT}: {P}re-training of
  deep bidirectional transformers for language understanding,'' in
  \emph{Proceedings of NAACL-HLT}, 2019, pp. 4171--4186.

\bibitem{MindInsight}
``The {MindInsight} repository on github,''
  \url{https://github.com/mindspore-ai/mindinsight}, Accessed: 2020-12-16.

\bibitem{xu2018ecglens}
K.~Xu, S.~Guo, N.~Cao, D.~Gotz, A.~Xu, H.~Qu, Z.~Yao, and Y.~Chen, ``{ECGLens}:
  {I}nteractive visual exploration of large scale {ECG} data for arrhythmia
  detection,'' in \emph{Proceedings of ACM SIGCHI}, 2018, p. 663.

\bibitem{ehrlinger2019daql}
L.~Ehrlinger, V.~Haunschmid, D.~Palazzini, and C.~Lettner, ``A {DaQL} to
  monitor data quality in machine learning applications,'' in \emph{Proceedings
  of International Conference on Database and Expert Systems Applications},
  2019, pp. 227--237.

\bibitem{du2018towards}
M.~Du, N.~Liu, Q.~Song, and X.~Hu, ``Towards explanation of {DNN}-based
  prediction with guided feature inversion,'' in \emph{Proceedings of ACM
  SIGKDD}, 2018, pp. 1358--1367.

\bibitem{kulesza2015principles}
T.~Kulesza, M.~Burnett, W.-K. Wong, and S.~Stumpf, ``Principles of explanatory
  debugging to personalize interactive machine learning,'' in \emph{Proceedings
  of International Conference on Intelligent User Interfaces}, 2015, pp.
  126--137.

\bibitem{wongsuphasawat2017visualizing}
K.~Wongsuphasawat, D.~Smilkov, J.~Wexler, J.~Wilson, D.~Mane, D.~Fritz,
  D.~Krishnan, F.~B. Vi{\'e}gas, and M.~Wattenberg, ``Visualizing dataflow
  graphs of deep learning models in {TensorFlow},'' \emph{IEEE Transactions on
  Visualization and Computer Graphics}, vol.~24, no.~1, pp. 1--12, 2017.

\bibitem{VisualDL}
``The {VisualDL} repository on github,''
  \url{https://github.com/PaddlePaddle/VisualDL}, Accessed: 2020-12-16.

\bibitem{DBLP:journals/tvcg/HohmanPRC20}
F.~Hohman, H.~Park, C.~Robinson, and D.~H.~P. Chau, ``Summit: {S}caling deep
  learning interpretability by visualizing activation and attribution
  summarizations,'' \emph{IEEE Transactions on Visualization and Computer
  Graphics}, vol.~26, no.~1, pp. 1096--1106, 2020.

\bibitem{DBLP:journals/tvcg/LiuSLLZL17}
M.~{Liu}, J.~{Shi}, Z.~{Li}, C.~{Li}, J.~{Zhu}, and S.~{Liu}, ``Towards better
  analysis of deep convolutional neural networks,'' \emph{IEEE Transactions on
  Visualization and Computer Graphics}, vol.~23, no.~1, pp. 91--100, 2017.

\bibitem{DBLP:conf/ieeevast/MingCZLCSQ17}
Y.~Ming, S.~Cao, R.~Zhang, Z.~Li, Y.~Chen, Y.~Song, and H.~Qu, ``Understanding
  hidden memories of recurrent neural networks,'' in \emph{Proceedings of IEEE
  Conference on Visual Analytics Science and Technology}, 2017, pp. 13--24.

\bibitem{DBLP:journals/tvcg/StrobeltGPR18}
H.~Strobelt, S.~Gehrmann, H.~Pfister, and A.~M. Rush, ``{LSTMVis}: {A} tool for
  visual analysis of hidden state dynamics in recurrent neural networks,''
  \emph{IEEE Transactions on Visualization and Computer Graphics}, vol.~24,
  no.~1, pp. 667--676, 2018.

\bibitem{DBLP:journals/tvcg/WangGYS18}
J.~Wang, L.~Gou, H.~Yang, and H.~Shen, ``{GANViz}: {A} visual analytics
  approach to understand the adversarial game,'' \emph{IEEE Transactions on
  Visualization and Computer Graphics}, vol.~24, no.~6, pp. 1905--1917, 2018.

\bibitem{larman2012applying}
C.~Larman, ``Applying {UML} and patterns: {An} introduction to object oriented
  analysis and design and interative development,'' 2012.

\bibitem{samek2017explainable}
W.~Samek, T.~Wiegand, and K.-R. Müller, ``{Explainable Artificial
  Intelligence}: {U}nderstanding, visualizing and interpreting deep learning
  models,'' \emph{ITU Journal: ICT Discoveries}, vol.~1, no.~1, pp. 39--48,
  2017.

\bibitem{liu2017analyzing}
M.~Liu, J.~Shi, K.~Cao, J.~Zhu, and S.~Liu, ``Analyzing the training processes
  of deep generative models,'' \emph{IEEE Transactions on Visualization and
  Computer Graphics}, vol.~24, no.~1, pp. 77--87, 2017.

\bibitem{kahng2017cti}
M.~Kahng, P.~Y. Andrews, A.~Kalro, and D.~H.~P. Chau, ``{Activis}: {V}isual
  exploration of industry-scale deep neural network models,'' \emph{IEEE
  Transactions on Visualization and Computer Graphics}, vol.~24, no.~1, pp.
  88--97, 2017.

\bibitem{bauerle2021net2vis}
A.~B{\"a}uerle, C.~Van~Onzenoodt, and T.~Ropinski, ``Net2vis--a visual grammar
  for automatically generating publication-tailored cnn architecture
  visualizations,'' \emph{IEEE transactions on visualization and computer
  graphics}, vol.~27, no.~6, pp. 2980--2991, 2021.

\bibitem{chollet2015keras}
F.~Chollet \emph{et~al.}, ``Keras,'' \url{https://keras.io}, 2015.

\bibitem{draw_convnet}
``The {draw-convnet} repository on github,''
  \url{https://github.com/gwding/draw_convnet}, Accessed: 2022-08-25.

\bibitem{convnet_drawer}
``The {convnet-drawer} repository on github,''
  \url{https://github.com/yu4u/convnet-drawer}, Accessed: 2022-08-25.

\bibitem{krizhevsky2012alexnet}
A.~Krizhevsky, I.~Sutskever, and G.~E. Hinton, ``{ImageNet} classification with
  deep convolutional neural networks,'' in \emph{Proceedings of Advances in
  neural information processing systems}, 2012, pp. 1097--1105.

\bibitem{lecun1998lenet}
Y.~LeCun, L.~Bottou, Y.~Bengio, and P.~Haffner, ``Gradient-based learning
  applied to document recognition,'' \emph{Proceedings of the IEEE}, vol.~86,
  no.~11, pp. 2278--2324, 1998.

\bibitem{sandler2018mobilenetv2}
M.~Sandler, A.~Howard, M.~Zhu, A.~Zhmoginov, and L.-C. Chen, ``{MobileNetV2}:
  {I}nverted residuals and linear bottlenecks,'' in \emph{Proceedings of IEEE
  CVPR}, 2018, pp. 4510--4520.

\bibitem{he2016resnet}
K.~He, X.~Zhang, S.~Ren, and J.~Sun, ``Deep residual learning for image
  recognition,'' in \emph{Proceedings of IEEE CVPR}, 2016, pp. 770--778.

\bibitem{jia2019optimizing}
Z.~Jia, J.~J. Thomas, T.~Warszawski, M.~Gao, M.~Zaharia, and A.~Aiken,
  ``Optimizing {DNN} computation with relaxed graph substitutions,'' in
  \emph{Proceedings of Machine Learning and Systems}, 2019, pp. 1--13.

\bibitem{MindSpore}
``The {MindSpore} repository on github,''
  \url{https://github.com/mindspore-ai/mindspore}, Accessed: 2020-12-16.

\bibitem{gunning2019darpa}
D.~Gunning and D.~W. Aha, ``{DARPA}'s explainable artificial intelligence
  program,'' \emph{AI Magazine}, vol.~40, no.~2, pp. 44--58, 2019.

\bibitem{LIU201748}
S.~Liu, X.~Wang, M.~Liu, and J.~Zhu, ``Towards better analysis of machine
  learning models: {A} visual analytics perspective,'' \emph{Visual
  Informatics}, vol.~1, no.~1, pp. 48--56, 2017.

\bibitem{chatzimparmpas2020state}
A.~Chatzimparmpas, R.~M. Martins, I.~Jusufi, K.~Kucher, F.~Rossi, and
  A.~Kerren, ``The state of the art in enhancing trust in machine learning
  models with the use of visualizations,'' \emph{Computer Graphics Forum},
  vol.~39, no.~3, pp. 713--756, 2020.

\bibitem{Du2019Tech}
D.~Mengnan, N.~Liu, and X.~Hu, ``Techniques for interpretable machine
  learning,'' \emph{Communications of the ACM}, vol.~63, no.~1, pp. 68--77,
  2019.

\bibitem{DBLP:journals/cvm/YuanCYLXL21}
J.~Yuan, C.~Chen, W.~Yang, M.~Liu, J.~Xia, and S.~Liu, ``A survey of visual
  analytics techniques for machine learning,'' \emph{Computational Visual
  Media}, vol.~7, no.~1, pp. 3--36, 2021.

\bibitem{DBLP:journals/tvcg/KrausePB14}
J.~Krause, A.~Perer, and E.~Bertini, ``{INFUSE:} {I}nteractive feature
  selection for predictive modeling of high dimensional data,'' \emph{IEEE
  Transactions on Visualization and Computer Graphics}, vol.~20, no.~12, pp.
  1614--1623, 2014.

\bibitem{DBLP:journals/tvcg/BernardHZFS18}
J.~Bernard, M.~Hutter, M.~Zeppelzauer, D.~W. Fellner, and M.~Sedlmair,
  ``Comparing visual-interactive labeling with active learning: {An}
  experimental study,'' \emph{IEEE Transactions on Visualization and Computer
  Graphics}, vol.~24, no.~1, pp. 298--308, 2018.

\bibitem{DBLP:journals/tvcg/AhnL20}
Y.~Ahn and Y.~Lin, ``{FairSight:} {V}isual analytics for fairness in decision
  making,'' \emph{IEEE Transactions on Visualization and Computer Graphics},
  vol.~26, no.~1, pp. 1086--1095, 2020.

\bibitem{DBLP:journals/tvcg/DingenVHMKBW19}
D.~Dingen, M.~van~'t Veer, P.~Houthuizen, E.~H.~J. Mestrom, H.~H.~M. Korsten,
  A.~R.~A. Bouwman, and J.~J. van Wijk, ``{RegressionExplorer:} {I}nteractive
  exploration of logistic regression models with subgroup analysis,''
  \emph{IEEE Transactions on Visualization and Computer Graphics}, vol.~25,
  no.~1, pp. 246--255, 2019.

\bibitem{DBLP:journals/tvcg/BergerMS17}
M.~Berger, K.~McDonough, and L.~M. Seversky, ``{C}ite2vec: {C}itation-driven
  document exploration via word embeddings,'' \emph{IEEE Transactions on
  Visualization and Computer Graphics}, vol.~23, no.~1, pp. 691--700, 2017.

\bibitem{DBLP:journals/tvcg/AndrienkoAABBFH21}
G.~L. Andrienko, N.~V. Andrienko, G.~Anzer, P.~Bauer, G.~Budziak, G.~Fuchs,
  D.~Hecker, H.~Weber, and S.~Wrobel, ``Constructing spaces and times for
  tactical analysis in football,'' \emph{IEEE Transactions on Visualization and
  Computer Graphics}, vol.~27, no.~4, pp. 2280--2297, 2021.

\bibitem{1532820}
F.-Y. Tzeng and K.-L. Ma, ``Opening the black box-data driven visualization of
  neural networks,'' in \emph{Proceedings of IEEE Visualization}, 2005, pp.
  383--390.

\bibitem{DBLP:journals/corr/Owhadi2022}
H.~Owhadi, ``Computational graph completion,'' \emph{Research in the
  Mathematical Sciences}, vol.~9, no.~2, pp. 27--59, 2022.

\bibitem{NETRON}
``The {NETRON} repository on github,''
  \url{https://github.com/lutzroeder/netron}, Accessed: 2021-09-17.

\bibitem{HiddenLayer}
``The {HiddenLayer} repository on github,''
  \url{https://github.com/waleedka/hiddenlayer}, Accessed: 2020-12-16.

\bibitem{sugiyama1981methods}
K.~Sugiyama, S.~Tagawa, and M.~Toda, ``Methods for visual understanding of
  hierarchical system structures,'' \emph{IEEE Transactions on Systems, Man,
  and Cybernetics}, vol.~11, no.~2, pp. 109--125, 1981.

\bibitem{6312901}
C.~{Batini}, E.~{Nardelli}, and R.~{Tamassia}, ``A layout algorithm for data
  flow diagrams,'' \emph{IEEE Transactions on Software Engineering}, vol.~12,
  no.~4, pp. 538--546, 1986.

\bibitem{Eschbach06orthogonalhypergraph}
T.~Eschbach, W.~Günther, and B.~Becker, ``Orthogonal hypergraph drawing for
  improved visibility,'' \emph{Journal of Graph Algorithms and Applications},
  vol.~10, no.~2, pp. 141--157, 2006.

\bibitem{eades1996orthogonal}
P.~Eades and Q.-W. Feng, ``Orthogonal grid drawing of clustered graphs,'' in
  \emph{Proceedings of Graph Drawing}, 1996, pp. 1--10.

\bibitem{DBLP:conf/gd/EadesF96}
P.~Eades and Q.~Feng, ``Multilevel visualization of clustered graphs,'' in
  \emph{Proceedings of Graph Drawing}, 1996, pp. 101--112.

\bibitem{221135}
E.~R. Gansner, E.~Koutsofios, S.~C. North, and K.-P. Vo, ``A technique for
  drawing directed graphs,'' \emph{IEEE Transactions on Software Engineering},
  vol.~19, no.~3, pp. 214--230, 1993.

\bibitem{10.1007/3-540-58950-3_371}
G.~Sander, ``Graph layout through the {VCG} tool,'' in \emph{Proceedings of
  Graph Drawing}, 1995, pp. 194--205.

\bibitem{10.1007/978-3-642-11805-0_14}
M.~Sp{\"o}nemann, H.~Fuhrmann, R.~von Hanxleden, and P.~Mutzel, ``Port
  constraints in hierarchical layout of data flow diagrams,'' in
  \emph{Proceedings of Graph Drawing}, 2010, pp. 135--146.

\bibitem{DBLP:journals/csur/CockburnKB08}
A.~Cockburn, A.~K. Karlson, and B.~B. Bederson, ``A review of overview+detail,
  zooming, and focus+context interfaces,'' \emph{ACM Computing Surveys},
  vol.~41, no.~1, pp. 1--31, 2008.

\bibitem{DBLP:journals/tochi/SchafferZGBDDR96}
D.~Schaffer, Z.~Zuo, S.~Greenberg, L.~Bartram, J.~Dill, S.~Dubs, and
  M.~Roseman, ``Navigating hierarchically clustered networks through fisheye
  and full-zoom methods,'' \emph{ACM Transactions on Computer-Human
  Interaction}, vol.~3, no.~2, pp. 162--188, 1996.

\bibitem{von2011visual}
T.~Von~Landesberger, A.~Kuijper, T.~Schreck, J.~Kohlhammer, J.~J. van Wijk,
  J.-D. Fekete, and D.~W. Fellner, ``Visual analysis of large graphs:
  {S}tate-of-the-art and future research challenges,'' in \emph{Proceedings of
  Computer Graphics Forum}, 2011, pp. 1719--1749.

\bibitem{DBLP:conf/apvis/BalzerD07}
M.~Balzer and O.~Deussen, ``Level-of-detail visualization of clustered graph
  layouts,'' in \emph{Proceedings of {APVIS}}, 2007, pp. 133--140.

\bibitem{DBLP:journals/vlc/HuangL06}
X.~Huang and W.~Lai, ``Clustering graphs for visualization via node
  similarities,'' \emph{Journal of Visual Languages and Computing}, vol.~17,
  no.~3, pp. 225--253, 2006.

\bibitem{DBLP:conf/gd/QuigleyE00}
A.~J. Quigley and P.~Eades, ``{FADE:} {G}raph drawing, clustering, and visual
  abstraction,'' in \emph{Proceedings of Graph Drawing}, 2000, pp. 197--210.

\bibitem{larochelle2009exploring}
H.~Larochelle, Y.~Bengio, J.~Louradour, and P.~Lamblin, ``Exploring strategies
  for training deep neural networks.'' \emph{Journal of machine learning
  research}, vol.~10, no.~1, pp. 1--40, 2009.

\bibitem{DBLP:journals/tvcg/ArchambaultMA08}
D.~Archambault, T.~Munzner, and D.~Auber, ``{GrouseFlocks}: {S}teerable
  exploration of graph hierarchy space,'' \emph{IEEE Transactions on
  Visualization and Computer Graphics}, vol.~14, no.~4, pp. 900--913, 2008.

\bibitem{abadi2016tensorflow}
M.~Abadi, P.~Barham, J.~Chen, Z.~Chen, A.~Davis, J.~Dean, M.~Devin,
  S.~Ghemawat, G.~Irving, M.~Isard, M.~Kudlur, J.~Levenberg, R.~Monga,
  S.~Moore, D.~G. Murray, B.~Steiner, P.~Tucker, V.~Vasudevan, P.~Warden,
  M.~Wicke, Y.~Yu, and X.~Zheng, ``{TensorFlow}: {A} system for large-scale
  machine learning,'' in \emph{Proceedings of USENIX Conference on Operating
  Systems Design and Implementation}, 2016, pp. 265--283.

\bibitem{paszke2019pytorch}
A.~Paszke, S.~Gross, F.~Massa, A.~Lerer, J.~Bradbury, G.~Chanan, T.~Killeen,
  Z.~Lin, N.~Gimelshein, L.~Antiga \emph{et~al.}, ``{Pytorch}: {An} imperative
  style, high-performance deep learning library,'' in \emph{Proceedings of
  Advances in neural information processing systems}, 2019, pp. 8026--8037.

\bibitem{holten2006hierarchical}
D.~Holten, ``Hierarchical edge bundles: {V}isualization of adjacency relations
  in hierarchical data,'' \emph{IEEE Transactions on Visualization and Computer
  Graphics}, vol.~12, no.~5, pp. 741--748, 2006.

\bibitem{7192715}
M.~{Sun}, P.~{Mi}, C.~{North}, and N.~{Ramakrishnan}, ``{BiSet}: {S}emantic
  edge bundling with biclusters for sensemaking,'' \emph{IEEE Transactions on
  Visualization and Computer Graphics}, vol.~22, no.~1, pp. 310--319, 2016.

\bibitem{estebanez2014performance}
C.~Est{\'e}banez, Y.~Saez, G.~Recio, and P.~Isasi, ``Performance of the most
  common non-cryptographic hash functions,'' \emph{Software: Practice and
  Experience}, vol.~44, no.~6, pp. 681--698, 2014.

\bibitem{BostockOH11}
M.~Bostock, V.~Ogievetsky, and J.~Heer, ``D{\({^3}\)} data-driven documents,''
  \emph{IEEE Transactions on Visualization and Computer Graphics}, vol.~17,
  no.~12, pp. 2301--2309, 2011.

\bibitem{parisi2012webgl}
T.~Parisi, ``{WebGL}: {U}p and running,'' 2012.

\bibitem{DBLP:conf/eccv/LiuAESRFB16}
W.~Liu, D.~Anguelov, D.~Erhan, C.~Szegedy, S.~E. Reed, C.~Fu, and A.~C. Berg,
  ``{SSD:} {S}ingle shot multibox detector,'' in \emph{Proceedings of ECCV},
  2016, pp. 21--37.

\bibitem{DBLP:journals/corr/SimonyanZ14a}
K.~Simonyan and A.~Zisserman, ``Very deep convolutional networks for
  large-scale image recognition,'' in \emph{Proceedings of ICLR}, 2015, pp.
  1--14.

\bibitem{DBLP:conf/cvpr/SzegedyVISW16}
C.~Szegedy, V.~Vanhoucke, S.~Ioffe, J.~Shlens, and Z.~Wojna, ``Rethinking the
  inception architecture for computer vision,'' in \emph{Proceedings of IEEE
  CVPR}, 2016, pp. 2818--2826.

\bibitem{DBLP:conf/nips/ShiCWYWW15}
X.~Shi, Z.~Chen, H.~Wang, D.~Yeung, W.~Wong, and W.~Woo, ``Convolutional {LSTM}
  network: {A} machine learning approach for precipitation nowcasting,'' in
  \emph{Proceedings of NeurIPS}, 2015, pp. 802--810.

\end{thebibliography}
%



%
\newpage

\begin{IEEEbiography}[{\includegraphics[width=1in,height=1.25in,clip,keepaspectratio]{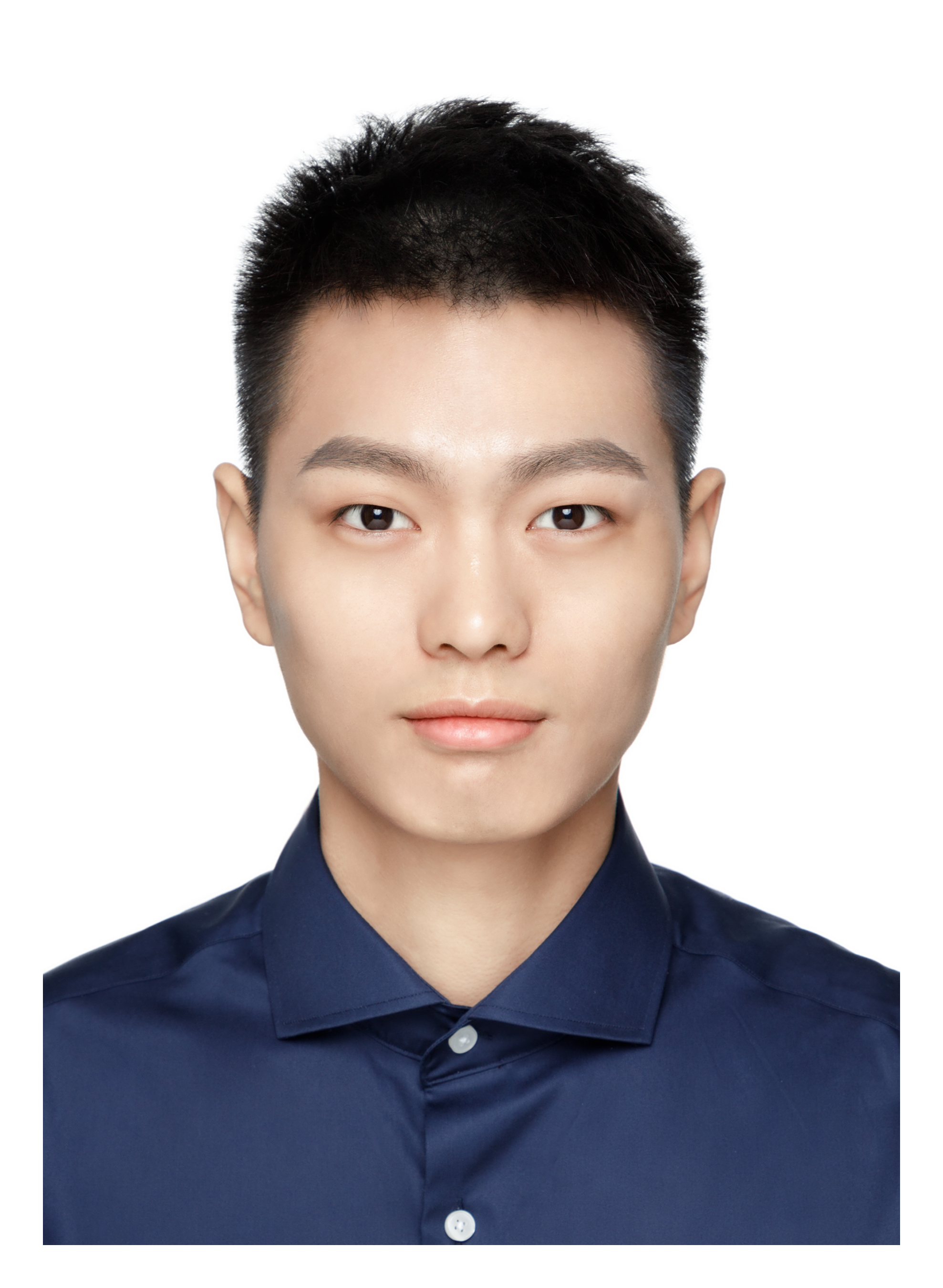}}]{Rusheng Pan} is a P.h.D. candidate in the State Key Lab of CAD\&CG at Zhejiang University, Hangzhou. He earned the B.S. degree in control science and engineering from Zhejiang University in 2018. His research interests are privacy preservation and graph visualization.
\end{IEEEbiography}

\begin{IEEEbiography}[{\includegraphics[width=1in,height=1.25in,clip,keepaspectratio]{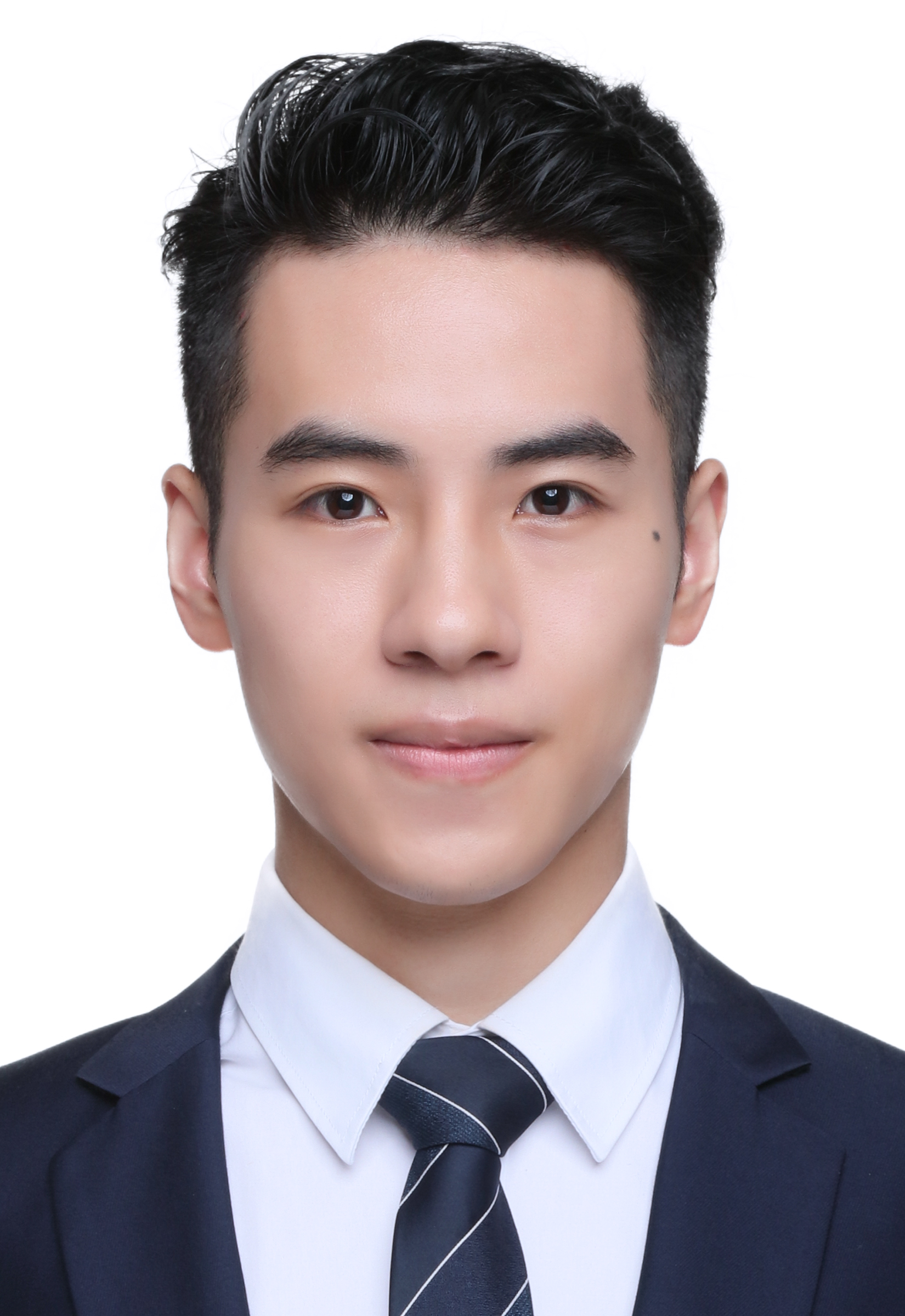}}]{Zhiyong Wang} is a postgraduate in the State Key Lab of CAD\&CG at Zhejiang University, Hangzhou. His research interests are visual analytics.
\end{IEEEbiography}

\begin{IEEEbiography}[{\includegraphics[width=1in,height=1.25in,clip,keepaspectratio]{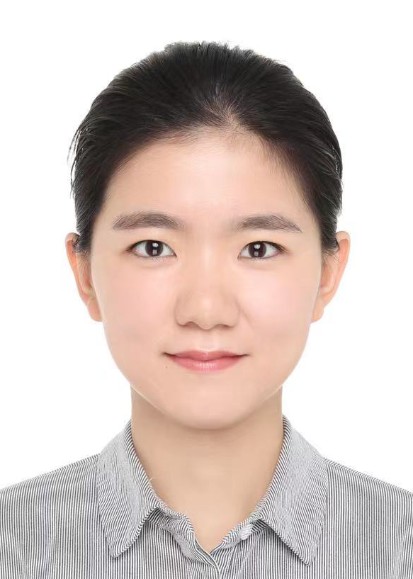}}]{Yating Wei} is a Ph.D. student in the State Key Lab of CAD\&CG at Zhejiang University, Hangzhou. She earned the B.S. degree in software engineering from Central South University in 2017. Her research interests are visual analytics and perceptual consistency.
\end{IEEEbiography}

\begin{IEEEbiography}[{\includegraphics[width=1in,height=1.25in,clip,keepaspectratio]{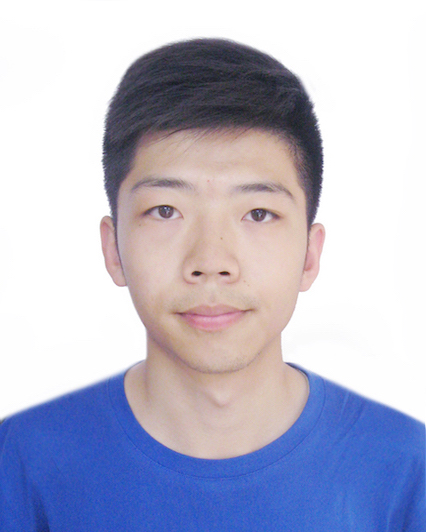}}]{Han Gao} received the MS degree from Big Data Technology Program, The Hong Kong University of Science and Technology, Hong Kong SAR, China, in 2019. He is now a research engineer with the Distributed Data Lab in CSI, Huawei Technologies Co., Ltd.
\end{IEEEbiography}

\begin{IEEEbiography}[{\includegraphics[width=1in,height=1.25in,clip,keepaspectratio]{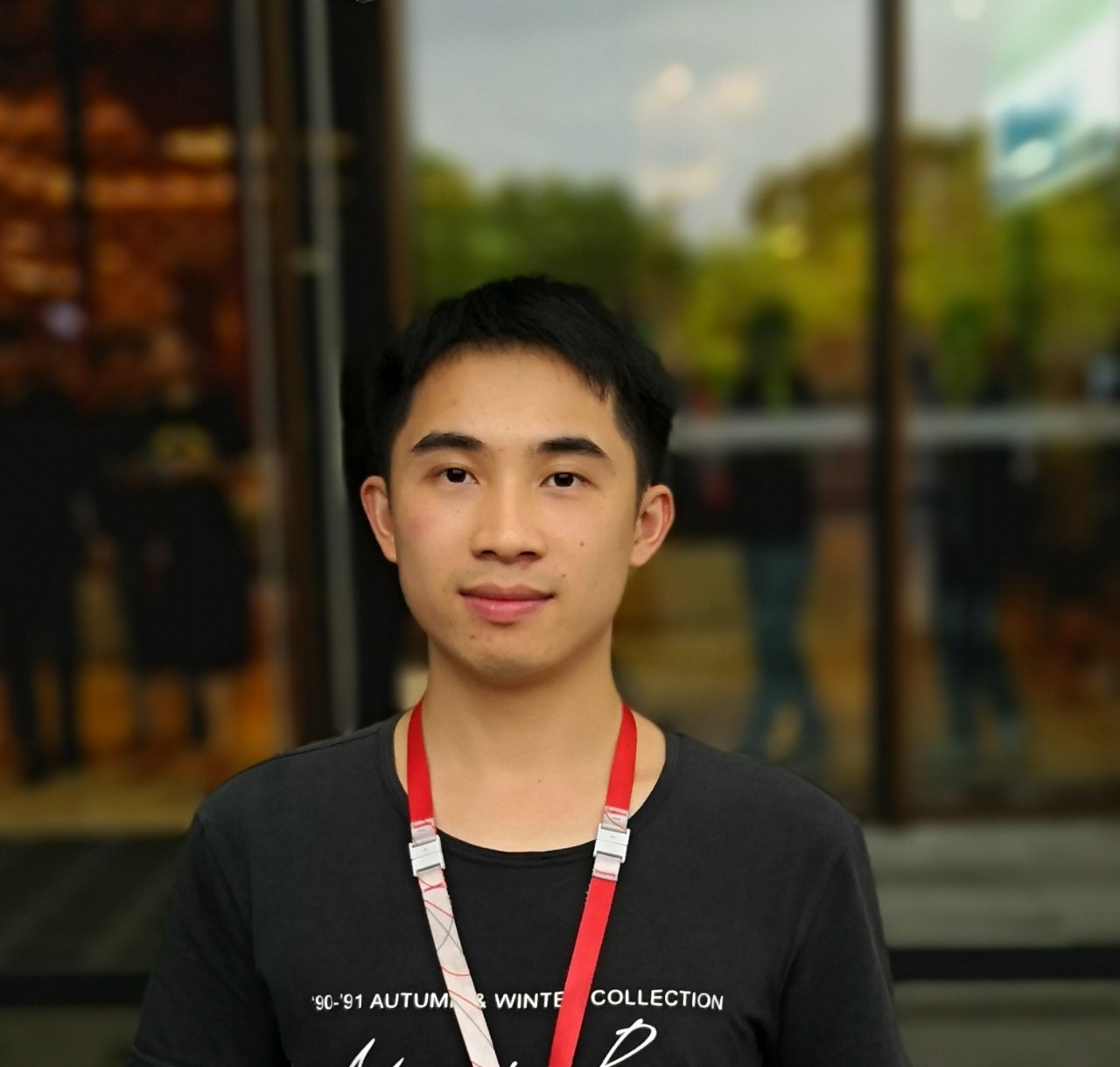}}]{Gongchang Ou} received the bachelor's degree from Chongqing University, in 2018. He is now a research engineer with the Distributed Data Lab in CSI, Huawei Technologies Co., Ltd.
\end{IEEEbiography}

\begin{IEEEbiography}[{\includegraphics[width=1in,height=1.25in,clip,keepaspectratio]{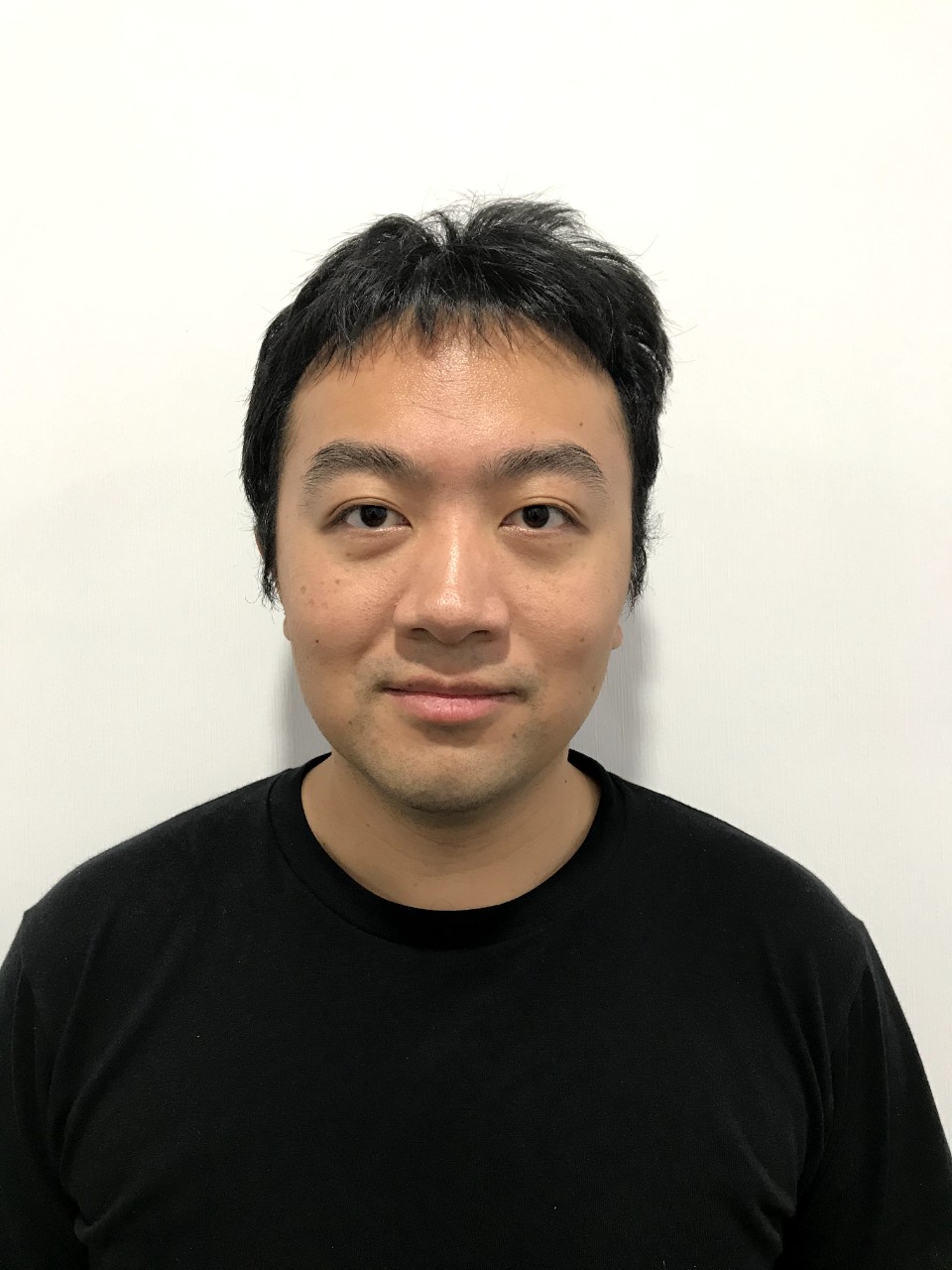}}]{Caleb Chen Cao} is a specialist in the Dist. Data Lab in CSI, Huawei. He received the Ph.D degree in Computer Science from HKUST. His research interests include explainable AI, AI governance and data fairness.
\end{IEEEbiography}

\begin{IEEEbiography}[{\includegraphics[width=1in,height=1.25in,clip,keepaspectratio]{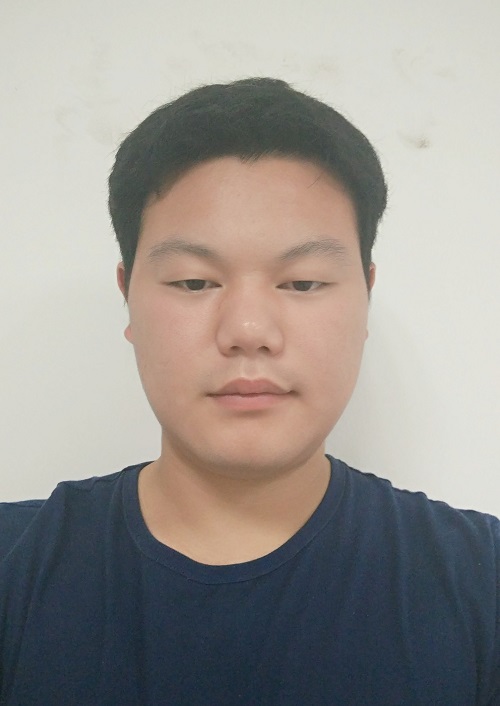}}]{Jingli Xu} is a postgraduate in the State Key Lab of CAD\&CG at Zhejiang University, HangZhou. He earned the B.S. degree in computer science and technology from Nanjing Tech University. His research interests are visualization and visual analytics.
\end{IEEEbiography}

\begin{IEEEbiography}[{\includegraphics[width=1in,height=1.25in,clip,keepaspectratio]{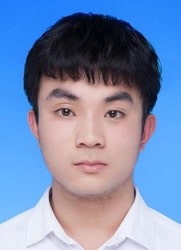}}]{Tong Xu} is a postgraduate in the State Key Lab of CAD\&CG at Zhejiang University, Hangzhou. His research interests are visual analytics and information visualization.
\end{IEEEbiography}

\begin{IEEEbiography}[{\includegraphics[width=1in,height=1.25in,clip,keepaspectratio]{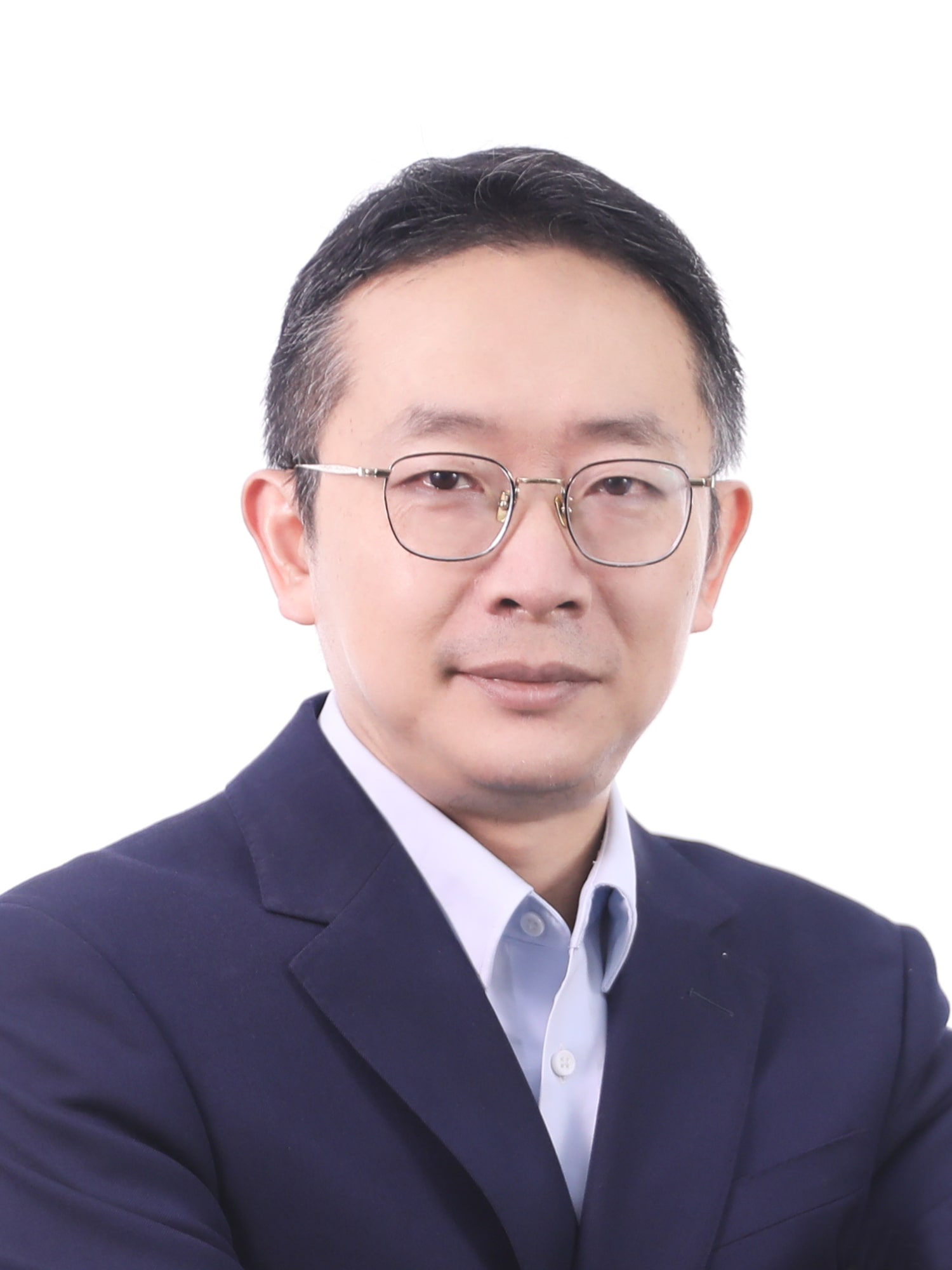}}]{Wei Chen} is a professor in the State Key Lab of CAD\&CG, Zhejiang University. His research interests include visualization and visual analysis, and has published more than 80 IEEE/ACM Transactions and IEEE VIS papers. He actively served as several associate editors of ACM/IEEE Transactions. 
\end{IEEEbiography}






\end{document}